\documentclass[aps,prb,twocolumn,superscriptaddress]{revtex4}
\usepackage[pdftex]{graphicx}
\usepackage{bm}
\usepackage{pifont}
\usepackage{cancel}
\usepackage{amssymb}
\usepackage{colordvi}
\usepackage{color}

\def\aa#1{\textcolor{black}{#1}}
\def\YU#1{\textcolor{black}{#1}}

\usepackage{soul,xcolor}
\setstcolor{red}

\usepackage{ae}
\usepackage[T1]{fontenc}
\usepackage[ansinew]{inputenc}
\usepackage{amsmath}

\usepackage[colorlinks]{hyperref}
\usepackage{epstopdf}

\def\sig{{\mbox{\boldmath{$\sigma$}}}}

\usepackage{soul,xcolor}
\setstcolor{red}

\begin{document}
\title{Spin selectivity through time-reversal symmetric helical junctions}
%
\author{Yasuhiro Utsumi}
\address{Department of Physics Engineering, Faculty of Engineering, Mie University, Tsu, Mie, 514-8507, Japan}

\author{Ora Entin-Wohlman}
\affiliation{School of Physics and Astronomy, Tel Aviv University, Tel Aviv 69978, Israel}

\author{Amnon Aharony}
\affiliation{School of Physics and Astronomy, Tel Aviv University, Tel Aviv 69978, Israel}


\begin{abstract}
Time-reversal symmetric charge and spin  transport through a molecule comprising two-orbital channels and connected to two leads is analyzed. It is demonstrated that spin-resolved currents are  generated when  spin-flip  processes are accompanied by a flip of the  orbital channels. This surprising finding does not contradict Bardarson's theorem [J. H. Bardarson, J. Phys. A: Math. Theor. {\bf 41}, 405203 (2008)] for two-terminal junctions: the transmission does possess two pairs of doubly-degenerate  eigenvalues as required by the theorem. The spin-filtering effect is  explicitly demonstrated   for a two-terminal chiral molecular junction,  modeled by a two-orbital  tight-binding chain with intra-atomic spin-orbit  interactions (SOI). In the context of transport through organic molecules   like DNA, this effect  is termed ``chirality-induced spin selectivity" (CISS). The model exhibits spin-splitting without breaking time-reversal symmetry: the intra-atomic SOI induces concomitant spin and orbital flips. Examining these transitions from the point of view of the Bloch states in an infinite molecule, it is shown that they cause shifts in the Bloch wave numbers, of the size of the reciprocal single turn, whose directions depend on the left-and right-handedness of the helix. As a result, spin-up and -down states propagate in the opposite directions, leading to the CISS effect. To further substantiate our picture, we present  an analytically-tractable expression for the 8$\times$8 scattering matrix of such a (single) molecule.
 \end{abstract}

\date{\today}
\maketitle


\section{Introduction}
\label{Intro}

Unexpectedly, a large spin-filtering effect has been observed in chiral molecules
~\cite{Goehler2011,Xie2011,Naaman2019}: injected electrons  become spin polarized after  being transmitted through a DNA molecule.
This  effect has been called  the ``chirality-induced spin selectivity" (CISS) ~\cite{Naaman2012,Michaeli2016,Michaeli2017}.
It is  a remarkable effect  since the organic molecules do not contain magnetic atoms, which would be apparent candidates for inducing spin-dependent phenomena.
Early theoretical attempts to explain this phenomenon attributed the preferential transmission
of electrons polarized along the same direction as the sense of advance of the helical molecule \cite{Yeganeh2009} or to the combination of  a weak Rashba interaction \cite{Rashba}  with weakly-dispersive electronic bands \cite{Gutierrez2012}.
An early theoretical paper~\cite{Guo2012} claimed that the CISS effect  results from the interplay between the spin-orbit interaction (SOI), the double-helix structure,  and the dissipation induced by leakage currents.
However, it was later pointed out that in the presence of long-range tunneling amplitudes connecting the atoms on the molecule  it suffices to consider just leakage currents from a single-stranded helix in order to produce the effect~\cite{Guo2014,Matityahu2016}.
Other theoretical papers discussed CISS by considering electron transport through a  double-helical pathway~\cite{Gutierrez2013}, in a double-stranded DNA~\cite{SierraBioMol2020}, and in a helical-tube geometry~\cite{Michaeli2019,Geyer2020}. These  papers are largely based on the linear-response regime of the transport,
though the model proposed in Ref. \onlinecite{Medina2015} relies on the possibility of the bias across the junction to select a specific spin orientation.
Recently, the possible significance of  transport in the nonlinear regime was arguably considered~\cite{Yang2019}, followed by a proposal for detecting the chirality from magnetoresistance measurements~\cite{Yang2019a}.

Most of these theoretical models~\cite{Gutierrez2012,Guo2012,Gutierrez2013,SierraBioMol2020,Guo2014,Geyer2020,Matityahu2016,Matityahu2017,Michaeli2019} exploit the Rashba-type SOI, which acts on bonds between atoms on the chain.
Then, in order to obtain an amount of spin filtering comparable with the  experimentally-detected values one needs to invoke a rather strong Rashba SOI. A strong inter-atomic SOI can be achieved when the $\sigma$ and $\pi$ orbitals on neighboring atoms are mixed due to the curved geometry ~\cite{HuertasHernandoPRB2006,VarelaPRB2016}. An example is the  intra-atomic SOI, of the order of $10$  meV $\sim$ 100 K for carbon atoms~\cite{Serrano2000,Kuemmeth2008}, which induces transitions from a $\pi $ orbital to a $\sigma$ orbital.
In such a case the curved geometry allows for electron hopping to the $\pi$ orbital of a nearest neighbor \cite{HuertasHernandoPRB2006}. These processes induce perturbatively an effective inter-atomic SOI proportional to the intra-atomic one \cite{DM}.
It was also suggested that the electric fields associated  with the hydrogen bonds of the base pairs can enhance the Rashba SOI by as much as a few tens meV~\cite{Varela2019}.
\aa{Such models, in which the effective Rashba SOI acts on electron transfer between $\pi$ orbitals of DNA bases~\cite{VarelaPRB2016,Varela2019}, were recently claimed to yield spin selectivity~\cite{VarelaPRB2020}.}
Recently, it was also proposed that the geometry-dependent relativistic origin of the SOI can be of order 100 meV in a nanoscale helix~\cite{Shitade2020}.

A major constraint on spin-resolved transport between two terminals arises from Bardarson's theorem~\cite{Bardarson2008}: for the simplest case of a single-channel junction, spin selectivity  through two-terminal time-reversal symmetric systems is forbidden.
Several papers proposed ways to overcome this constraint in that simple case.
One such way is to break time-reversal symmetry by including magnetic fields~\cite{AAA}.
Magnetic fields  applied on the reservoirs connected to the junction may also produce spin selectivity~\cite{Aharony2019}.
However, since many of the experiments on the CISS effect did not include magnetic fields, it is desirable to find theoretical ways which generate spin filtering without breaking time-reversal symmetry.
This can be achieved  by  utilizing  junctions  connected to three (or more) terminals~\cite{Pareek2004,Yokoyama2009,Matityahu2017}.
An example of that is  the leak currents mentioned above~\cite{Guo2012,Guo2014,Matityahu2016}, which are accounted for by connecting  more  leads, i.e., B\"uttiker probes
~\cite{Buettiker1988}.

In this paper we follow an alternative route, in which the spin-orbit interaction causes scattering between  sub bands in the junction. This idea was introduced  in Ref.~\onlinecite{Eto2005} as a way to establish spin filtering  in
quantum point contacts (QPC's)~\cite{Scheid2010,Kohda2012}, in tubular two-dimensional gases~\cite{Entin-WohlmanPRB2010}, and in  quasi-one dimensional quantum wires~\cite{NagaevPRB2014}.
A common feature of  these setups is that they involve  more than two channels, each of which comprises up- and down-spin channels, e.g.,  the first and second sub bands of the QPC.
In the context of  the CISS theory, Refs.~\onlinecite{Gutierrez2013,Michaeli2019,SierraBioMol2020,Geyer2020,Medina2015} may fall into this category.

A widely-spread  belief  in the CISS community is that the two-terminal system cannot exhibit the CISS effect.
One possible reason  for this might arise from an extended interpretation of Bardarson's theorem \cite{Bardarson2008}.
Bardarson showed that {\it in time-reversal-symmetric systems with half integer spins, the transmission eigenvalues of the scattering matrix come in degenerate pairs}.
Assuming that this Kramers-type degeneracy involves spins with opposite eigenvalues, the theorem prohibits the two-terminal spin filtering because each pair of doubly-degenerate transmission eigenvalues carries the same amount of up and down spins.
However, the theorem does not specify which spin states are associated with the doubly-degenerate transmission eigenvalues.
Therefore, it is possible to consider, e.g., two pairs of doubly-degenerate transmission eigenvalues in which one pair carries two up spins in one direction and the other pair carries two down spins in the opposite direction.
Hence, the theorem does not rule out the `counter examples'~\cite{Scheid2010,Kohda2012,Entin-WohlmanPRB2010,NagaevPRB2014} of the no-go theorem of   spin filtering by  two-terminal  setups.

In this paper we present a detailed analysis of a spin filter that consists of a two-orbital molecule (in total four channels  when spin indices are included),  which is connected to two terminals.
\YU{Like most of the earlier work on two-terminal systems, we consider only elastic scattering, which must obey the Bardarson theorem.}
The corresponding  (8$\times$8) scattering  matrix, in which spin and channel flips occur simultaneously in a way that preserves time-reversal symmetry is analysed and the two doubly-degenerate transmission eigenvalues are identified.
 Importantly, each such an eigenvalue corresponds to two spin states which are not oppositely directed, and therefore we do obtain a  spin-polarized conductance through a two-terminal, time-reversal symmetric  junction.

Following this general idea, we introduce a specific example of a toy model describing a single strand of a double-stranded DNA: a two-orbital helical tight-binding chain with an intra-atomic SOI.
Our model possesses two advantages:
(1) It does not require leaky leads, and thus is close to the experimental setups
~\cite{Goehler2011,Xie2011};
(2)  The amount of spin-splitting achieved in this model is comparable to the bare intra-atomic SOI { coupling,  multiplied by the `normalized' curvature of the helix.
As opposed, the magnitude of the Rashba-type interaction \cite{Rashba} due to the mixing  of the $\sigma$ and $\pi$ orbitals,   which results from the curved geometry as discussed above,  is smaller than the bare intra-atomic SOI,  by more than a factor of $100$ according to Ref.~\onlinecite{HuertasHernandoPRB2006}.
We use our model to demonstrate numerically the spin filtering in a two-terminal setup.

Our findings are substantiated in two ways: First we explore the band structure of an infinite chain of such molecules and identify certain features in it that are related to the possibility of spin-resolved transport. Second,  we examine analytically a simplified version of our model and show that the spin polarization calculated within it complies with the numerical results.

The structure of the paper is as follows. Section \ref{sec2}
begins with a review of  Bardarson's theorem and an analysis of an $8 \times 8$ two-orbital  scattering matrix (Sec.~\ref{sec:Bardarson_T}). Section \ref{sec:conductance}  continues with the derivation
of the two-terminal spin and charge conductances and presents   a discussion of the conditions required to realize  spin-filtering.
The latter is shown in Sec. \ref{sec:spin_filter} to be finite for the two-orbital junction. Section \ref{QPC} (augmented by Appendix
\ref{sec:effective_hamiltonian_QPC})
demonstrates that our picture pertains also to spin-filtering in a quantum point contact in which the Rashba interaction is active.
We introduce in Sec.~\ref{sec:model_tohtbm}  the Hamiltonian of our two-orbital helical tight-binding chain with intra-atomic SOI (with more details given in Appendix \ref{sec:effective_hamiltonian}),  and in  Sec. \ref{sec:S_matrix_spin_current} we derive the corresponding scattering matrix. These are used in Sec. \ref{sec:numerical_result} to compute numerically the spin polarization of our model. As mentioned, the band structure
of the infinite helix, calculated in Sec. ~\ref{sec:band_structure},  allows one to explore  certain features related to the CISS effect. Section ~\ref{sec:Analytical_S_matrix} is separated into two parts. Section \ref{syms} discusses certain symmetry properties of the scattering matrix of the two-channel junction, while Sec. \ref{analyticex} presents an analytical solution of the 8$\times$8 scattering matrix of a simplified version of our toy model, which yields
analytically-tractable exact results for the spin polarization (details are given in Appendix \ref{Tech}).
We believe that the lack of  such tractable expressions  has left  doubts in the CISS community concerning  the possibility to obtain spin-filtering   in a two-terminal setup obeying time-reversal symmetry. Our results are summarized in Sec.~\ref{sec:summary}.
Throughout the paper we set $\hbar=e=1$, and use the terms `orbital' and `channel' alternatively; those do not include the spin degrees of freedom.


\section{General theory of spin filtering in a two-terminal two-orbital junction}
\label{sec2}

\subsection{Bardarson's theorem}
\label{sec:Bardarson_T}

We begin our discussion with a short summary of  Ref.~\onlinecite{Bardarson2008}; this will also serve to introduce our   notations.
According to  Bardarson \cite{Bardarson2008}, there are two ways to represent the time-reversal operation. In one approach~\cite{JJSakurai},  the time-reversal operator $\Theta$ changes the wave  incoming from lead $s$  with wave vector ${\bf k}$ and spin index $\sigma$,  $ |{\bf k} ; \alpha \sigma \rangle_s$,  into the outgoing wave $|-{\bf k} ; \alpha \bar{\sigma} \rangle_s$,
\begin{align}
\Theta |{\bf k} ; \alpha \sigma \rangle_s = \sigma |-{\bf k} ; \alpha \bar{\sigma} \rangle_s\ ,
\label{tro}
\end{align}
where $\bar{\sigma}=\downarrow (\uparrow)$ for $\sigma=\uparrow (\downarrow)$  and $\alpha$ is the channel (orbital) index.
When $\sigma$ appears as a coefficient it should be read as $\sigma = +1 (-1)$ for $\sigma=\uparrow (\downarrow)$. Hence, the scattering state in the left ($s=L$) terminal or in the  right ($s=R$) one becomes
\begin{align}
|\psi \rangle^{}_s = |{\bf k};\alpha \sigma \rangle^{}_s c_s^{\rm in}  + \left( \Theta |{\bf k} ; \alpha \sigma\rangle^{}_s \right) \bar{c}_s^{\rm out} \ .
\label{eqn:scattering_state}
\end{align}
The amplitudes $c^{\rm in}_{s}$ and $\overline{c}^{\rm out}_{s}$ (the latter is  the time-reversed partner of the former) are defined as follows. For $N^{}_s$ channels in terminal $s$, the ket vector consists of 2$N^{}_s$ components,
\begin{align}
|{\bf k} \rangle^{}_s =\left[ |{\bf k} ; 1 \uparrow \rangle^{}_s,  |{\bf k} ; 1 \downarrow \rangle^{}_s,
\cdots,  |{\bf k} ; N_s \uparrow \rangle^{}_s, |{\bf k} ; N^{}_s \downarrow \rangle^{}_s \right] \, .
\end{align}
Correspondingly, the amplitudes $c^{\rm in}_{s}$ and $\overline{c}^{\rm out}_{s}$  are 2$N^{}_s-$component  vectors,
\begin{align}
c_s^{{\rm in}} = \left[ \begin{array}{c} c_{1 \uparrow s}^{{\rm in} } \\ c_{1 \downarrow s}^{{\rm in} } \\ \vdots \\ c_{N_s \uparrow s}^{{\rm in} } \\ c_{N_s \downarrow s}^{{\rm in} }
\end{array} \right]\ ,
\;\;\;\;
\bar{c}_s^{{\rm out}} = \left[ \begin{array}{c} \bar{c}_{1 \downarrow s}^{{\rm out} } \\ \bar{c}_{1 \uparrow s}^{{\rm out}} \\ \vdots \\  \bar{c}_{N_s \downarrow s}^{{\rm out} } \\ \bar{c}_{N_s \uparrow s}^{{\rm out}} \end{array} \right] \, .
\label{amp}
\end{align}

The scattering matrix  connects the coefficients of the  incoming and outgoing waves,
\begin{align}
\bar{c}^{\rm out}_{}= \bar{S} c^{\rm in}_{} \, ,
\;\;\;\;
\bar{c}^{\rm out} _{}= \left[ \begin{array}{c} \bar{c}_L^{\rm out} \\ \bar{c}_R^{\rm out} \end{array} \right] \, ,
\;\;\;\;
c^{\rm in}_{}= \left[ \begin{array}{c} c_L^{\rm in} \\ c_R^{\rm in} \end{array} \right] \, , \label{eqn:s_bar_F}
\end{align}
where
\begin{align}
\bar{S} = \left[ \begin{array}{cc} \bar{r} & \bar{t}' \\ \bar{t} & \bar{r}' \end{array} \right] \ .
\label{Sbar}
\end{align}
The `bar' notation on the scattering-matrix entries indicates the time-reversal operation, Eq. (\ref{tro}).
When $N^{}_s=N^{}_L=N_R^{}$, each sub matrix in Eq. (\ref{Sbar}) is of order 2$N^{}_{s} \times$2$N^{}_{s}$.
The unitarity relation $\bar{S} \bar{S}^\dagger=\bar{S}^\dagger \bar{S}={\bm 1}_{4 N^{}_{s}}$
(where ${\bf 1}_{4N^{}_{s}}$ is the  4$N^{}_{s} \times$4$N^{}_{s}$ unit matrix}),
ensures the relations
\begin{align}
{\bar{r}} {\bar{r}}^\dagger + {\bar{t'}} {\bar{t'}}^\dagger = {\bar{r}}^\dagger_{} {\bar{r}} + {\bar{t}}^\dagger_{} {\bar{t}} = {\bm 1}_{2 N^{}_{s}} \, .
\end{align}
The time-reversal of the scattering state in Eq.  (\ref{eqn:scattering_state}) is
\begin{align}
\Theta |\psi \rangle^{}_s = \left( \Theta |{\bf k};\alpha \sigma \rangle^{}_s \right) (c_s^{\rm in})^*  - |{\bf k};\alpha \sigma \rangle^{}_s (\bar{c}_s^{\rm out})^* \ ,
\label{nss}
\end{align}
where the property  $\Theta^2=-1$, and  the fact that $\Theta$ includes  the complex-conjugation operation, have been used. As seen, the new scattering state
(\ref{nss}) is just
the original scattering state (\ref{eqn:scattering_state}), with the replacements
$c_s^{\rm in} \to -(\bar{c}_s^{\rm out})^*$
and
$\bar{c}_s^{\rm out} \to  ({c}_s^{\rm in})^*$.
It therefore follows from Eq. (\ref{eqn:s_bar_F}) that
\begin{align}
(c^{\rm in})^* = \bar{S} (- \bar{c}^{\rm out})^* \, . \label{eqn:s_bar_B}
\end{align}
Then, comparing Eqs.~(\ref{eqn:s_bar_F}) and (\ref{eqn:s_bar_B})
and exploiting the unitarity of the scattering matrix $\bar{S}$, 
Bardarson~\cite{Bardarson2008} concluded that the scattering matrix is antisymmetric. This implies that the reflection-amplitude matrix is also antisymmetric, and hence  can be represented as~\cite{Schliemann2001} $\bar{r} = V {\rm diag}(i \lambda^{}_1 \sigma^{}_y,\cdots, i \lambda_{N^{}_{s}} \sigma^{}_y) V^T_{}$, where $V$ is a unitary matrix, and the superscript $T$ indicates a transposed matrix.
It follows that
$\bar{r}^\dagger \bar{r} = V {\rm diag}(\lambda_1^2 \sigma^{}_0,\cdots, \lambda^{2}_{N^{}_{s}}\sigma^{}_0) V_{}^T$ and thus the transmission eigenvalues, i.e., the eigenvalues $\lambda^{2}_\alpha$ of $\bar{t}^\dagger \bar{t} = \bm{1}^{}_{2 N^{}_{s}} -\bar{r}^\dagger \bar{r}$ come in degenerate pairs~\cite{Bardarson2008}. Here, $\sigma^{}_0={\bf 1}^{}_2$ and $\sigma^{}_y$ is the usual Pauli matrix.

Alternatively, the ubiquitous way of implementing the time-reversal operation,  which is  also used in the following, defines the scattering state as
\begin{align}
|\psi \rangle^{}_s = |{\bf k} \rangle^{}_s c_s^{\rm in}  + |-{\bf k} \rangle^{}_s c_s^{\rm out}  \, ,
\label{eqn:scattering_state_}
\end{align}
where the scattering matrix $S$ comprises the time-reversed partners of the entries of $\bar{S}$, Eq. (\ref{Sbar}),
\begin{align}
{S} = \left[\begin{array}{cc} {r} & {t}' \\ {t} & {r}' \end{array} \right] \, .
\label{eqn:s_matrix}
\end{align}
In this basis
\begin{align}
{c}^{\rm out} = {S} c^{\rm in} \, ,
\;\;\;\;
{c}^{\rm out} = \left[ \begin{array}{c} {c}_L^{\rm out} \\ {c}_R^{\rm out} \end{array} \right] \, ,
\;\;\;\;
c^{\rm in}= \left[\begin{array}{c} c_L^{\rm in} \\ c_R^{\rm in} \end{array} \right]\, .
\label{eqn:s_def_1}
\end{align}
Since the time-reversed state and the outgoing state are related by
$\Theta |{\bf k} \rangle^{}_s = |-{\bf k} \rangle^{}_s [{\bf 1}^{}_{2N} \otimes (i \sigma^{}_y)] $
a comparison of Eqs.~(\ref{eqn:scattering_state}) and (\ref{eqn:scattering_state_}) yields
\begin{align}
c_s^{\rm out} = {\bf 1}^{}_{2N} \otimes (i \sigma^{}_y) \bar{c}_s^{\rm out}\ ,
\end{align}
 (i.e. $c^{\rm out}_{\alpha \uparrow s} = \bar{c}^{\rm out}_{\alpha \uparrow s}$ and $c^{\rm out}_{\alpha \downarrow s} = - \bar{c}^{\rm out}_{\alpha \downarrow s}$) and thus the scattering matrix satisfies the relation $S= {\bf 1}_{2 N^{}_{s}} \otimes (i \sigma_y) \, \bar{S}$,
and  is self-dual~\cite{Bardarson2008},
\begin{align}
S=({\bf 1}_{2 N^{}_{s}} \otimes \sigma^{}_y) S^T ({\bf 1}_{2 N^{}_{s}} \otimes \sigma^{}_y) \ ,
\label{eqn:BT}
\end{align}
where $S^{T}_{}$ is the transposed scattering matrix.
The block-diagonal component of the scattering  matrix, i.e., the matrix of the reflection coefficients, satisfies
$r=({\bf 1}_{N^{}_{s}} \otimes \sigma^{}_y) r^T ({\bf 1}_{N^{}_{s}} \otimes \sigma^{}_y)$.
Hence, the reflection amplitude from  the state with orbital $\alpha'$ and spin index  $\sigma'$ into the state with orbital $\alpha$ and spin index $\sigma$, $r^{}_{\alpha \sigma,\alpha' \sigma'}$, is such that
\begin{align}
r^{}_{\alpha \sigma , \alpha' \sigma'} = \sigma \sigma' \, r^{}_{\alpha' \bar{\sigma}' , \alpha \bar{\sigma}} \, .
\label{eqn:sym_ref}
\end{align}
This relation is very useful for the following considerations.


\subsection{Charge and spin conductances}
\label{sec:conductance}

Here we present the definitions of the linear-response conductances for charge and spin flows, in terms of the scattering-matrix elements. Our derivation is specific for a two-terminal, two-channel junction, in which the spin degree of freedom is relevant, implying  an 8$\times$8 matrix.

Assign a chemical potential $\mu^{}_{s}$
to the $s$ terminal,
and denote  the charge and  spin currents into the $s$ terminal by $I^{}_{j;s}$, where
$j=0$ pertains to the charge current and $j=x, y, $ and $z$  to the three spin currents.
Then, the formal expressions for the conductances in terms of the scattering matrix $S$ are
\begin{align}
G^{}_{j;ss} &=\frac{\partial I^{}_{s;j}}{\partial \mu^{}_s}= \frac{1}{2\pi} {\rm Tr}  \{ (\Pi^{}_{ s } \otimes \tau^{}_0 \otimes \sigma^{}_{j}) \nonumber\\
&\times\left[{\bm 1}_8 - S (\Pi^{}_{ s } \otimes \tau^{}_0 \otimes \sigma^{}_0) S^\dagger _{}\right] \} \ ,
\label{eqn:g_for_analytic1}
\end{align}
and
\begin{align}
&G^{}_{j; s \bar{s}}=- \frac{\partial I^{}_{s;j}}{\partial \mu^{}_{\bar{s}}} \nonumber\\
&= \frac{1}{2\pi} {\rm Tr} \{(\Pi^{}_{ s } \otimes \tau^{}_0 \otimes \sigma^{}_j) S (\Pi^{}_{ \bar{s} } \otimes \tau^{}_0 \otimes \sigma^{}_0) S^\dagger_{} \}\ .
\label{eqn:g_for_analytic2}
\end{align}
Here,  $\Pi^{}_{L}={\rm diag}(1,0)$ and $\Pi^{}_{R}={\rm diag}(0,1)$ are matrices of projection operators.
The unit matrix in the orbital space is  ${\tau}^{}_0={\bm 1}^{}_2$, and $\sigma^{}_{j}$ is the $j$th Pauli matrix.

Exploiting the decomposition of the scattering matrix into sub matrices of transmission and reflection amplitudes, Eq.~(\ref{eqn:s_matrix}), one obtains
\begin{align}
G^{}_{j;LL} &=  {\rm Tr} \{ \tau^{}_0 \otimes \sigma^{}_j ( {\bm 1}^{}_4 - r r^\dagger _{}) \}/(2\pi)\ , \nonumber\\
G^{}_{j; LR} &= {\rm Tr} \{ \tau^{}_0 \otimes \sigma^{}_j {t'} {t'}^\dagger_{} \}/(2\pi) \ , \nonumber\\
G^{}_{j; RL} &={\rm Tr} \{\tau_0 \otimes \sigma_j {t} {t}^\dagger \}/(2\pi) \ , \nonumber\\
G^{}_{j;RR} &=  {\rm Tr} \{ \tau^{}_0 \otimes \sigma^{}_j ( {\bm 1}^{}_4 - r' {r'}^\dagger ) \}/(2\pi)
\ . \label{eqn:Gj_}
\end{align}
The self duality of the scattering matrix (\ref{eqn:BT}) leads to another peculiar feature of the spin conductances.
Whereas interchanging the order of the reflection (transmission) matrix and its Hermitian conjugate (for instance, $rr^{\dagger}_{}\rightarrow r^{\dagger}_{}r$) in the expressions in Eq. (\ref{eqn:Gj_}) does not change the charge conductances (for which $j=0$), it reverses the sign of the spin conductances, $j=x,y,z$.
By  the unitarity of the scattering matrix, $SS^\dagger={\bm 1}_8$, one easily verifies that
\begin{align}
G^{}_{j; ss} = G^{}_{j; s \bar{s}} \ ,
\label{eqn:symmetry}
\end{align}
which ensures that the net charge and  spin currents at equilibrium vanish.
Note that for the charge current, i.e., for $j=0$,
the unitarity of the scattering matrix  leads to
\begin{align}
G_{0; ss} = G_{0; \bar{s} s} \, ,
\end{align}
which  implies that the charge currents measured in the two leads are identical.

When the chemical potentials in the two leads  differ slightly, such that
$\mu^{}_{s}=\mu+\delta\mu^{}_{s}$ and
$\mu^{}_{\bar{s}}=\mu+\delta\mu^{}_{\bar{s}}$,
 the net spin or charge current flowing out of lead $s$ is
\begin{align}
I^{}_{j;s}(\delta \mu_s,\delta \mu_{ \bar{s}}) &=G^{}_{j; ss} \delta \mu^{}_s - G^{}_{j; s \bar{s}} \delta \mu^{}_{\bar{s}} \nonumber \\ &= G^{}_{j; ss} (\delta \mu^{}_s - \delta \mu^{}_{\bar{s}}) \ .
\label{Ijs}
\end{align}
Another consequence of Eqs. (\ref{eqn:symmetry}) and (\ref{Ijs})
is that the polarization direction of the spin current
is independent of the direction of the chemical potential bias,
\begin{align}
I^{}_{j;s}(\delta \mu,0) = -I^{}_{j;s}(0,\delta \mu) \ ,
\end{align}
which is a specific feature of the two-terminal setup.
The spin polarization factor, defined as
\begin{align}
P^{}_{j;s}= G^{}_{j;ss} /G^{}_{0;ss} = G^{}_{j;s \bar{s}} /G^{}_{0;s \bar{s}} \ ,
 \label{eqn:P}
\end{align}
is  also insensitive  to the direction of the chemical-potential bias (i.e., whether $\mu^{}_{s}>\mu^ {}_{\bar{s}}$ or else), as follows from Eq.~(\ref{eqn:symmetry}).
This feature  does not necessitate  time-reversal symmetry of the scatterer and is  independent of the number of  channels;
it applies to a two-terminal single-channel spin-filter. For example,  it was  found  in transport through a spin-orbit active weak-link in the presence of a  magnetic field  \cite{Shekhter2018}.


\subsection{Spin polarization}
\label{sec:spin_filter}

We first consider
the simplest configuration of a two-terminal junction with a single channel, for which $N_{s}^{}=1$, implying a 4$\times$4 scattering matrix.
Its 2$\times$2 reflection matrix is self-dual and consequently is  diagonal \cite{Bardarson2008}, as by Eq. (\ref{eqn:sym_ref})
$r^{}_{\uparrow , \uparrow} = r^{}_{\downarrow , \downarrow}=r^{}_0$
and
$r^{}_{\uparrow , \downarrow} = -r^{}_{\uparrow , \downarrow}=0$.
It follows that ${t}^\dagger {t} = \sigma^{}_0 - r^\dagger_{} r= (1-|r^{}_0|^2) \sigma^{}_0$
and thus the transmission eigenvalues are degenerate. As a result,
the spin conductance  and with it the spin polarization vanish,   $G_{j;ss}=0$ and $P_{j;s}=0$ ($j=x,y,z$).

Next we continue to the two-channel case where $N_{s}^{}=2$, which corresponds to an 8$\times$8 scattering matrix.
The amplitude vectors $c_{s}^{\rm in}$ and $c_{s}^{\rm out}$ given in Eqs.
(\ref{amp}) are then four-dimensional.
In the specific situation in which each reflection process changes or preserves both the spin and  channel indices, the corresponding reflection matrix takes the form
\begin{align}
r
&= \left[ \begin{array}{cccc}
r^{}_{1 \uparrow , 1 \uparrow} & 0^{}_{} & 0^{}_{} & r^{}_{1 \uparrow , 2 \downarrow} \\
0 & r^{}_{1 \downarrow , 1 \downarrow} & r^{}_{1 \downarrow , 2 \uparrow} & 0 \\
0 & r^{}_{2 \uparrow, 1 \downarrow} & r^{}_{2 \uparrow , 2 \uparrow} & 0 \\
r^{}_{2 \downarrow , 1 \uparrow} & 0 & 0 & r^{}_{2 \downarrow , 2 \downarrow}
\end{array} \right]\nonumber \\
&= \left[ \begin{array}{cccc}
r^{}_{1 \uparrow , 1 \uparrow} & 0^{}_{} & 0^{}_{} & r^{}_{1 \uparrow , 2 \downarrow} \\
0 & r^{}_{1 \uparrow , 1 \uparrow} & -r^{}_{2 \downarrow , 1 \uparrow} & 0 \\
0 & -r^{}_{1\uparrow,2\downarrow} & r^{}_{2 \downarrow , 2 \downarrow} & 0 \\
r^{}_{2 \downarrow , 1 \uparrow} & 0 & 0 & r^{}_{2 \downarrow , 2 \downarrow}
\end{array} \right] \ ,
\label{eqn:rm_2c}
\end{align}
where we have used the self-duality property (\ref{eqn:sym_ref}).
The matrix (\ref{eqn:rm_2c}) can be rearranged in a block-diagonal form,
$r={\rm diag}(r^{}_+,r^{}_-)$, where
\begin{align}
r^{}_+ = \left[ \begin{array}{cc} r^{}_{1 \uparrow , 1 \uparrow} & r^{}_{1 \uparrow , 2 \downarrow} \\ r^{}_{2 \downarrow , 1 \uparrow} & r^{}_{2 \downarrow , 2 \downarrow} \end{array} \right] ,\
r^{}_- = \left[ \begin{array}{cc} r^{}_{2 \downarrow , 2 \downarrow} & -r^{}_{1 \uparrow , 2 \downarrow} \\ -r^{}_{2 \downarrow , 1 \uparrow} & r^{}_{1 \uparrow , 1 \uparrow} \end{array} \right] \ . \label{eqn:rp_rm}
\end{align}
The two matrices  $r^{}_{+}$ and $r^{}_{-}$ are  time-reversed of one another,  $r^{}_- = \sigma^{}_y r_+^T \sigma^{}_y$.
The four transmission eigenvalues are the solutions of the  characteristic polynomial equation
\begin{align}
{\rm det} \{ \Lambda {\bm 1}_{4}- t^\dagger t \}
&= ( {\rm det} \{ (\Lambda-1) {\bm 1}^{}_{2} + r_\pm^\dagger r^{}_\pm \})^2=0 \ ,
\end{align}
and obviously come in pairs of degenerate eigenvalues,  $\Lambda_+$ and  $\Lambda_-$, implying that our model complies with Bardarson's theorem \cite{Bardarson2008}.
Explicitly, the degenerate eigenvalues are
\begin{align}
\Lambda^{}_\pm &= 1-X\pm \sqrt{X^2-|Y|^2} \ ,\nonumber\\
X &=[ |r^{}_{1 \uparrow , 1 \uparrow}|^2 + |r^{}_{1 \uparrow , 2 \downarrow}|^2 + |r^{}_{2 \downarrow , 1 \uparrow}|^2 + |r^{}_{2 \downarrow , 2 \downarrow}|^2]/2 \ ,\nonumber\\
Y &= r^{}_{1 \uparrow , 1 \uparrow} r^{}_{2 \downarrow , 2 \downarrow} - r^{}_{1 \uparrow , 2 \downarrow} r^{}_{2 \downarrow , 1 \uparrow} \ .
\label{ddt}
\end{align}
Inserting these results into the first  of Eqs.~(\ref{eqn:Gj_}) yields
\begin{align}
G^{}_{0;LL} &=(\Lambda^{}_{+}+\Lambda^{}_{-})/\pi\ ,\nonumber\\
G^{}_{z;LL} &=( |r^{}_{2 \downarrow, 1 \uparrow}|^2 - |r^{}_{1 \uparrow, 2 \downarrow}|^2 )/\pi\ ,\nonumber\\
G^{}_{x;LL} &=G^{}_{y;LL}=0 \ .
\label{GxyLL}
\end{align}
It therefore follows from Eq. (\ref{eqn:P})  that the spin polarization along $z$ is finite,
\begin{align}
P^{}_{z;L} = (|r_{2 \downarrow, 1 \uparrow}|^2 - |r_{1 \uparrow, 2 \downarrow}|^2 )/(\Lambda^{}_{+}+\Lambda^{}_{-}) \ .
\label{eqn:pzL}
\end{align}

The fact that our model for the reflection matrix, Eqs. (\ref{eqn:rm_2c}),  leads to spin polarization along $z$ alone, can be explained by inspecting the scattering states. The scattering state of
an electron with wave number $k$  and spin index  $\sigma$ injected  in channel $\alpha$ is
\begin{align}
 |\psi^{}_{\alpha \sigma} \rangle^{}_L  =  | k; \alpha \sigma \rangle^{}_L + |\delta \psi^{}_{\alpha \sigma} \rangle^{}_L \ ,
 \end{align}
  where the reflected wave, as dictated by  Eq.~(\ref{eqn:rm_2c}), is a superposition of states of different spins and channel indices,
\begin{align}
|\delta \psi^{}_{\alpha \sigma} \rangle^{}_L = |-k; \alpha \sigma \rangle^{}_L \, r^{}_{\alpha \sigma, \alpha \sigma} + |-k; \bar{\alpha} \bar{\sigma} \rangle^{}_L \, r^{}_{ \bar{\alpha} \bar{\sigma}, \alpha \sigma} \, ,
\end{align}
with $\bar{\alpha}=2(1)$ for $\alpha=1(2)$.
Since the spin  operator is diagonal in the channel index,
${}^{}_L \langle k ; \alpha \sigma| \hat{\sigma}^{}_j |k' ; \alpha' \sigma' \rangle^{}_L = \delta^{}_{k',k} \delta^{}_{\alpha',\alpha} \left( \sigma_j \right)^{}_{\sigma',\sigma}$,
the spin components for  $j=x$ and $j=y$ vanish. Only the charge conductance ($j=0$) and the spin-$z$  conductance ($j=z$) remain,
\begin{align}
G^{}_{j;LL} = \frac{1}{2 \pi} \sum_{\alpha,\gamma=\pm} {\rm Tr} \left[ \Pi^{}_\gamma \sigma^{}_j \Pi^{}_\gamma \left( \sigma^{}_0 - r^{}_\alpha r_\alpha^\dagger \right) \right]
\ , \label{eq:g_j_AharonyTheory}
\end{align}
where $\Pi^{}_+={\rm diag}(1,0)$ and $\Pi^{}_-={\rm diag}(0,1)$ are matrices of projection operators.

From Eq.~(\ref{eqn:pzL}), one can deduce the condition for a   perfect spin-filtering,
\begin{align}
P^{}_{z;L} =
\left \{ \begin{array}{cc}
1 &
(|r^{}_{2\downarrow,1\uparrow}|^{2}=1-(|r^{}_{1\uparrow,1\uparrow}|^{2}_{}+|r^{}_{2\downarrow,2\downarrow}|^{2})/2) \\
-1 &
(|r^{}_{1\uparrow,2\downarrow}|^{2}=1-(|r^{}_{1\uparrow,1\uparrow}|^{2}_{}+|r^{}_{2\downarrow,2\downarrow}|^{2})/2)
\end{array}
\right. \ .
\end{align}
This perfect polarization,  achieved in a   two-orbital,  two-terminal junction,   is in contrast with the locking of the directions of spin and momentum discussed  in Ref.~\onlinecite{Medina2015}, in which $\uparrow$ and $\downarrow$ spins propagate in  opposite directions.
In that case, the direction of the spin polarization changes when  the chemical potential bias is reversed. On the other hand, in our scenario  the direction of the spin polarization  in each lead is independent of the direction of the chemical potential bias as indicated by Eq.~(\ref{eqn:P}).

Since the matrices $r^{}_+$ and $r^{}_-$ are self dual to one another, they can be presented as quaternion numbers,
$r^{}_\pm = A^{}_0 \sigma^{}_0 \pm i {\bm A} \cdot {\bm \sigma}$.
In terms of these quaternions
\begin{align}
 r^{}_\pm r_\pm^\dagger& = |A^{}_0|^2+|{\rm Re} {\bm A}|^2 +|{\rm Im} {\bm A}|^2 \nonumber \\
&+ 2 (\pm {\rm Im} (A^{}_0 {\bm A})+ {\rm Re} {\bm A} \times {\rm Im} {\bm A} ) \cdot {\bm \sigma}
\ ,
\label{eqn:rr}
\end{align}
which implies that
although $r^{}_+$ and $r_-$ are dual to one  another, $ r^{}_{+} r_{+}^\dagger$ is not necessarily the dual of
$ r^{}_{-} r_{-}^\dagger$, that is,
$ r^{}_{-} r_{-}^\dagger= \sigma^{}_y (r_+^\dagger r_+^*)^T_{} \sigma^{}_y \neq \sigma^{}_y (r^{}_{+}r^{\dagger}_{+})^T_{} \sigma^{}_y
$. This implies that the directions of the two reflected states  are not necessarily opposite,
and opens the possibility to create a finite  conductance of the $z$ component of the spin,
\begin{align}
G^{}_{z;LL} = -\frac{8}{2 \pi}  \left( {\rm Re} {\bm A} \times {\rm Im} {\bm A} \right)^{}_z \ .
\end{align}
Combining this result with the charge conductance expressed in terms of the quaternions,
\begin{align}
G^{}_{0;LL} = \frac{8}{2 \pi}  \left( 1 -|A^{}_0|^2 - |{\rm Re} {\bm A}|^2-|{\rm Im} {\bm A}|^2 \right) \, ,
\end{align}
we find for the $z-$polarization factor Eq. (\ref{eqn:P})
\begin{align}
P^{}_{z;L} = -\frac{\left( {\rm Re} {\bm A} \times {\rm Im} {\bm A} \right)_z}{1 -|A^{}_0|^2 - |{\rm Re} {\bm A}|^2-|{\rm Im} {\bm A}|^2} \, .
\end{align}
This implies that to obtain the spin-filtering effect, $r^{}_\pm$ should be a complex quaternion number in the pseudo-spin space comprising the states $(\alpha,\sigma)$ and $(\overline{\alpha} ,\overline{\sigma})$.


\subsection{Filtering by  a   point contact subjected to the Rashba interaction}
\label{QPC}

From the above discussion it follows that mixing  an even number of channels by  spin-orbit interactions is crucial for realizing   spin filtering   without breaking time-reversal symmetry.
The importance of such mixing   was emphasized in previous papers, in which it was found that SOI-induced mixing of two sub bands in a quantum point contact (QPC) enables spin filtering
~\cite{Eto2005,Scheid2010}.
Here we demonstrate
that our two-terminal,  two-channel  scenario  applies also to a QPC subjected to the Rashba SOI. This adds a formal basis and further  insights for the findings of Ref. ~\onlinecite{Eto2005}.

\begin{figure}[ht]
\includegraphics[width=8.5cm]{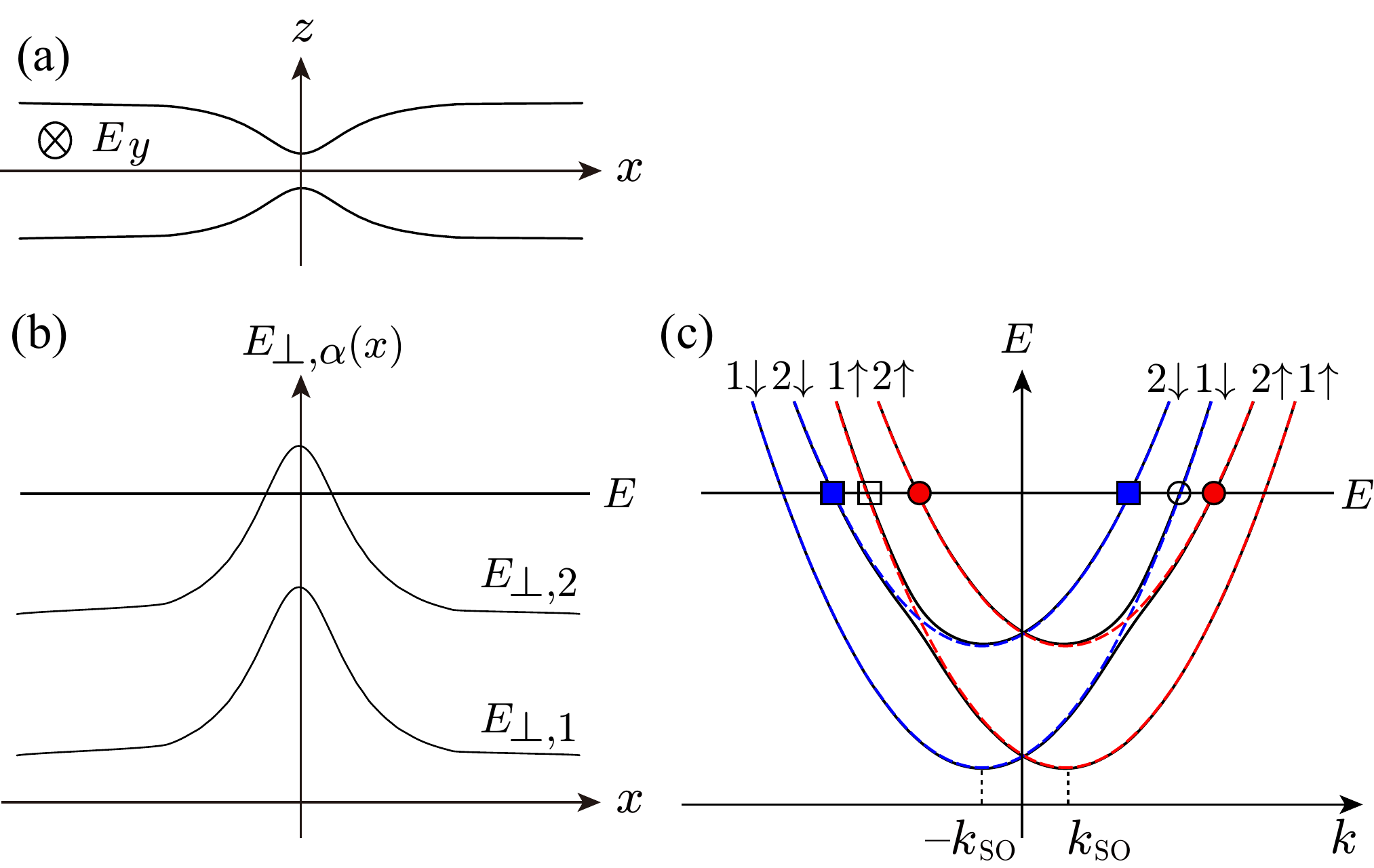}
\caption{(Color online) (a) Schematics  of a quantum point contact  subjected to the Rashba spin-orbit interaction caused by the electric field $E_{y}^{}$ along $y$.
(b) The effective potential induced by the spatial adiabatic change of the confining potential, as   represented by the  sub bands  $E_{\perp,1}^{}(x)$ and  $E_{\perp,2}^{}(x)$ (see text). (c) The dispersion relations of the first two  sub bands for electron traveling along $x$. 
The spin-orbit interaction  mixes the $1 \downarrow$ and  $2 \uparrow$ sub bands as well as the  $1 \uparrow$ and $2 \downarrow$ ones.
}
\label{fig:qpcsubband}
\end{figure}

Figure \ref{fig:qpcsubband}(a) depicts a schematic drawing of a QPC:
a two-dimensional electron gas  is confined to  the $x$-$z$ plane, and  subjected to a uniform electric field along $y$, $E^{}_{y}$, which appears since the confining potential lacks  the mirror symmetry $y \to -y$. This electric field  gives rise to the Rashba interaction \cite{Rashba},
${\cal H}^{}_{\rm SOR}= k^{}_{\rm so} (\sigma^{}_x\hat{p}^{}_{z}- \sigma^{}_z\hat{p}^{}_{x})/m^{}_{e}$, which couples the momentum ($\hat{\bf p}$) and the spin degrees of freedom. Here
$k^{}_{\rm so}\propto E^{}_{y}$
 characterizes the strength of the Rashba interaction, and $m^{}_{e}$ is the electron mass.

The confining potential along the $z$ direction,  $U(z;x)$,  is assumed to vary adiabatically as a function of $x$.
Under these conditions, one is able to construct a
quasi-one-dimensional Hamiltonian that describes the motion along $x$.
Consider first the  motion along $z$, at a fixed value of $x$, which is described by the Hamiltonian
${\cal H}^{}_\perp = \hat{p}_z^2/(2m^{}_{e}) + U(z;x)$, with eigen energies  $E_{\perp,\alpha}(x)$ and  orthonormal eigenfunctions $\chi^{}_\alpha(z;x)$, where $\alpha$ is the sub-band index.  The  energies $E_{\perp,1}(x)$ and $E_{\perp,2}(x)$ of the first two sub bands  are depicted in Fig.~\ref{fig:qpcsubband}(b).
They act as an adiabatic potential for the  motion in the  $x-$direction.

Setting the wave function of the entire Hamiltonian to be $\varphi(x,z)=\sum_{\alpha=1}^{\infty} \psi^{}_{\alpha}(x)\chi^{}_{\alpha}(z;x)$,
where $\psi^{}_{\alpha}(x)$ is a two-component spinor belonging to channel $\alpha$,
we derive the Schr\"odinger equation for the quasi-one-dimensional propagation along the $x-$direction,
$\sum_{\alpha^\prime=1}^\infty {\mathcal H}^{}_{\alpha,\alpha^\prime} \psi^{}_{\alpha^\prime}(x) = E \psi^{}_\alpha(x)$,
within the adiabatic approximation.  [Put differently,  the transverse wave function varies  very slowly along $x$, such that $\partial^{}_{x}\chi^{}_{\alpha}(z;x)\approx 0$ \cite{HeikkilaBook2013}.]
This effective quasi-one-dimensional Hamiltonian is
(see Appendix \ref{sec:effective_hamiltonian_QPC})
\begin{align}
{\cal H}^{}_{\alpha,\alpha'}\approx\left [ \frac{ (\hat{p}^{}_{x} - k^{}_{\rm so} \sigma^{}_z)^2}{2m^{}_{e}} +E^{}_{\perp,\alpha}(x)  \right  ] \delta^{}_{\alpha,\alpha'} + V^{}_{\alpha,\alpha'} \sigma^{}_x \ .
\label{eqn:Hq1d}
\end{align}
[A constant energy shift, $k^{2}_{\rm so}/(2m^{}_{e})$,  was omitted.]
The spin-flip mixing between sub
bands $\alpha$ and $\alpha'$ is caused by  the change of the transverse wave  function in the $z$ direction,
\begin{align}
V^{}_{\alpha,\alpha'} = \frac{\ k^{}_{\rm so}}{m^{}_{e}} \int dz \chi^{\ast}_{\alpha}(z;x) \hat{p}^{}_z  \chi^{}_{\alpha'}(z;x) \ .
\label{Vf}
\end{align}
The dispersion relations  of the  first two sub bands (for propagation along $x$ far from the constriction region around $x\sim0$) are portrayed in Fig.~\ref{fig:qpcsubband}(c); they are split by $\pm k^{}_{\rm so}$ for the two $z$ components of the spin.

In the absence of
$V_{\alpha,\alpha'}$, an electron injected
 from $x=-\infty$ in  sub band  2 with energy $E$ such that
$E_{\perp,2}(0)>E$  will be reflected backwards to $x=-\infty$, since in the constriction region its energy is not enough to traverse the potential barrier formed by $E_{\perp,2}(0)$
[see Fig.~\ref{fig:qpcsubband}(b)].
On the other hand, an electron in sub band  $1$  with energy
$E>E_{\perp,1}(x)$ is transmitted to $x=\infty$ without being reflected.
However, the Rashba SOI in Eq. (\ref{eqn:Hq1d}), aside from splitting the bands according to the spin indices by $\pm k^{}_{\rm so}$, leads also to avoided crossings between the $1 \downarrow$ and $2 \uparrow$ sub bands,
and between the $1 \uparrow$ and $2 \downarrow$ ones,  as shown in Fig.~\ref{fig:qpcsubband}(c): the term
$V^{}_{\alpha,\alpha'} \sigma^{}_x$ can flip both the spin and the sub band indices,  causing scattering between the $1 \downarrow(\uparrow)$ and $2 \uparrow(\downarrow)$ sub bands.
Eventually, during the scattering process, two right-going states, $2 \uparrow$ and $1 \downarrow$, and one left-going state $2 \uparrow$ [the filled circles and the empty circle in Fig.~\ref{fig:qpcsubband}(c)] can be mixed.
In the same way, their time-reversed partners, two left-going states, $2 \downarrow$ and $1 \uparrow$, and one right-going state $2 \downarrow$ [the filled squares and the empty square in Fig.~\ref{fig:qpcsubband}(c)] are mixed. Neglecting the remaining scattering processes, the reflection matrix
 that expresses these possibilities is
\begin{align}
r \approx \left[ \begin{array}{cccc}
0 & 0 & 0 & r^{}_{1 \uparrow , 2 \downarrow} \\
0 & 0 & 0 & 0 \\
0 & -r^{}_{1 \uparrow , 2 \downarrow} & r^{}_{2 \downarrow , 2 \downarrow} & 0 \\
0 & 0 & 0 & r^{}_{2 \downarrow , 2 \downarrow}
\end{array} \right] \ .
\end{align}
The doubly-degenerate transmission eigenvalues, Eqs. (\ref{ddt}),   of this matrix are
$\Lambda^{}_+=0$ and
$\Lambda^{}_-=1-|r^{}_{1 \uparrow , 2 \downarrow}|^2-|r^{}_{2 \downarrow , 2 \downarrow}|^2$, leading to a
polarization factor (\ref{eqn:pzL}) for the $z$ component of the spin
\begin{align}
P^{}_{z;L} \approx \frac{- |r^{}_{1 \uparrow, 2 \downarrow}|^2 }{2 - |r^{}_{2 \downarrow, 2 \downarrow}|^2  - |r^{}_{1 \uparrow, 2 \downarrow}|^2 } \ .
\end{align}
In the same way,  the matrix of reflection amplitudes for  an electron impinging from $x=\infty$ is
\begin{align}
r' \approx \left[ \begin{array}{cccc}
0 & 0 & 0 & 0 \\
0 & 0 & -r'^{}_{2 \downarrow , 1 \uparrow} & 0 \\
0 & 0 & r'^{}_{2 \downarrow , 2 \downarrow}& 0 \\
r'^{}_{2 \downarrow , 1 \uparrow} & 0 & 0 & r'^{}_{2 \downarrow , 2 \downarrow}
\end{array} \right] \ ,
\end{align}
with the  transmission eigenvalues are $\Lambda^{}_+=0$ and
$\Lambda^{}_-=1-|r^{\prime}_{2 \downarrow, 1 \uparrow}|^2-|r^{\prime}_{2 \downarrow , 2 \downarrow}|^2$.
In this case the polarization factor (\ref{eqn:pzL}) for spins along $z$ is
\begin{align}
P^{}_{z;R} \approx \frac{|r_{2 \downarrow, 1 \uparrow}'|^2 }{2 - |r_{2 \downarrow, 2 \downarrow}'|^2  - |r_{2 \downarrow, 1 \uparrow}'|^2 } \, .
\end{align}
For a system possessing  a mirror symmetry with respect to the  $y$-$z$ plane, one  expects that
$r'^{}_{2 \downarrow, 1 \uparrow}=r^{}_{2 \uparrow, 1 \downarrow}=-r^{}_{1 \uparrow, 2 \downarrow}$
and
$r'^{}_{2 \downarrow, 2 \downarrow}=r^{}_{2 \uparrow, 2 \uparrow}=r^{}_{2 \downarrow, 2 \downarrow}$ [see Eq. (\ref{eqn:r_reflectionsymmetric})], which implies
that  the polarization factors observed in the left and right lead  are opposite,
\begin{align}
P^{}_{z;R} = -P^{}_{z;L} \, .
\end{align}
In the absence of this
symmetry,  one may imagine that upon reflecting the system through that
plane the direction of the electric field is reversed, $E_{y}^{}\rightarrow -E^{}_{y}$. Were this field the only source of the spin-orbit coupling, then $k_{\rm so}$ will reverse its sign as well.
This reflection effectively reverses the direction of spin, $\sigma^{}_x \to -\sigma^{}_x$ and $\sigma^{}_z \to -\sigma^{}_z$ in the Hamiltonian (\ref{eqn:Hq1d}), and also reverses the sign of $V^{}_{\alpha,\alpha^\prime}$, Eq. (\ref{Vf}).
The symmetry of the scattering matrix [see Eq.~(\ref{eqn:trS_qpc})] then implies that the sign of the spin polarization factor is reversed.
This property is reminiscent of the one predicted  for the CISS effect (see, e.g., Refs.~\onlinecite{Naaman2012},\onlinecite{Guo2012}, \onlinecite{Matityahu2016}, and \onlinecite{Medina2015}): the interchange of left- and right-handedness reverses the sign of spin polarization.


\section{Spin-filtering through a two-terminal junction}
\label{secIII}


\subsection{Two-orbital  tight-binding Hamiltonian of a helical chain with intra-atomic spin-orbit interaction}
\label{sec:model_tohtbm}



\begin{figure}[ht]
\includegraphics[width=0.7 \columnwidth]{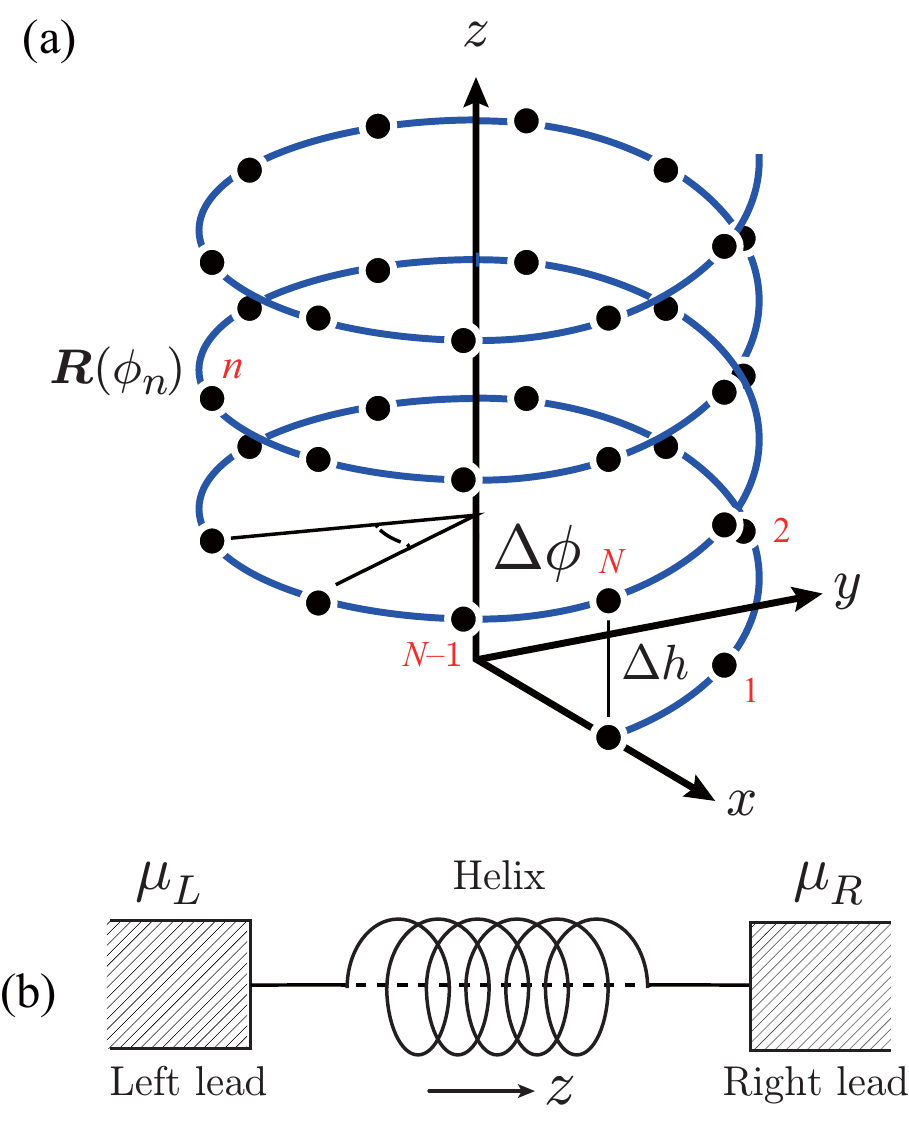}
\caption{
(Color online) (a) Schematic picture of a single strand of a double-stranded DNA. ${\bm R}(\phi^{}_{n})$ is the radius-vector of site $n$ within the
Frenet-Serret scheme [Eq. (\ref{Rphi})],   $\Delta h$ is the pitch,  $\Delta\phi=2\pi/N$, and $\phi^{}_{n}=n\Delta\phi$.
The original tight-binding Hamiltonian
(\ref{eqn:original_hamiltonian}) is expressed in the coordinate system $x$, $y$, and $z$, shown in the figure.
(b) A molecular junction; The left and right leads are attached to two edges of the single strand of the DNA molecule.
A difference in the chemical potentials of the left and right leads, $\mu^{}_{L}$ and $\mu^{}_{R}$, induces a flow of electrons. }
\label{fig:setupDNA}
\end{figure}

We exemplify the general discussion given in Sec. \ref{sec2} by studying a toy model: a  single strand of a double-stranded DNA molecule [Fig. \ref{fig:setupDNA}(a)], coupled to two leads [Fig. \ref{fig:setupDNA}(b)].  The molecule is represented by a helical  tight-binding chain, where  each atom accommodates three   $p-$orbitals with intra-atomic spin-orbit interaction.  This interaction is assumed to be strongly anisotropic, such that  the  $p^{}_{y}$ orbital [lying along the tangential direction of the thick curved line in Fig. \ref{fig:setupDNA}(a)]
is not accessible, and only the $p_{x}$ and  $p_{z}$ orbitals participate in the electron dynamics. This restriction renders the sites in our tight-binding chain  to be occupied  only by the orbitals $|p^{}_x \rangle = |x \rangle$ and $|p^{}_z \rangle = |z \rangle$. The  construction of the Hamiltonian of the molecule is detailed in
Appendix \ref{sec:effective_hamiltonian}, where we show that it takes the form
\begin{align}
{\cal H}^{}_{\rm mol} &= \sum_{n=1}^{N^{}_{\rm mol}-1} (-J c_{n+1}^\dagger c^{}_{n} + {\rm H.c.}) + \sum_{n=1}^{N^{}_{\rm mol}} \epsilon^{}_0 c_{n}^\dagger c^{}_{n} \nonumber \\ & + \Delta \epsilon c_{n}^\dagger \tau^{}_z \otimes  \sigma^{}_0 c^{}_{n}  + \Delta_{\rm so} c_{n}^\dagger  \tau^{}_y \otimes {\bm t}(\phi^{}_n) \cdot {\bm \sigma} c^{}_{n}  \ ,
\label{eq:H_mol}
\end{align}
where $N_{\rm mol}$ is the number of sites on the molecule. The creation operator on site $n$
\begin{align}
c^{\dagger}_{n} = \left[ \begin{array}{cccc} c^{\dagger}_{n ; x \uparrow} & c^{\dagger}_{n ; x \downarrow} & c^{\dagger}_{n ; z \uparrow}  &c^{\dagger}_{n ; z \downarrow} \end{array} \right] \ ,
\label{eqn:spinor_twoorb}
\end{align}
has four components as required for a two-orbital description that includes the spin degree of freedom. The first three terms on the right-hand side of Eq. (\ref{eq:H_mol}) are those of a standard tight-binding model, where $J$ is the tunneling amplitude between nearest-neighbor sites (assumed for simplicity to be identical for the two orbital and  spin indices), $\epsilon^{}_{0}$ is the on-site energy, and $\Delta\epsilon$
is the energy difference between the $p^{}_x$ and $p^{}_z$ orbitals. We assume that the unit cell of the helical molecule contains $N$ sites; the location of the $n$th site is specified by   $\phi_n=2 \pi n /N$ [see Fig. \ref{fig:setupDNA}(a) and
Appendix \ref{sec:effective_hamiltonian}].

The key ingredient of the  Hamiltonian is the fourth term on the right-hand side of Eq. (\ref{eq:H_mol}), which describes the intra-atomic spin-orbit coupling, of strength $\Delta^{}_{\rm so}$ ($\tau_{j}$, with $j=0,x,y,z$,  are the Pauli matrices in the orbital space comprising $p_{x}$ and $p_{z}$). As seen, this term is proportional to the inner product of the Pauli matrix-vector ${\bm \sigma}=\{\sigma^{}_x,\sigma^{}_y,\sigma^{}_z\}$,
and the tangent vector along the spiral axis of the molecule,  ${\bf t}(\phi)$. [The term `spiral axis' refers to the thick curved line in Fig. \ref{fig:setupDNA}(a).] Written in terms of the radius $R$ of the helix and  its pitch $\Delta h$, the tangent vector is
\begin{align}
{\bf t}(\phi) =L  \{- \tilde{\kappa} \sin (\phi), p \tilde{\kappa} \cos (\phi), |\tilde{\tau}|\} \ ,
\label{tphi}
\end{align}
where    $L=  \sqrt{R^2+[\Delta h/(2 \pi)]^2}$,  and     $p$ specifies
the chirality of helix: $p=1 (-1)$ for a right-handed (left-handed) helix~\cite{Matityahu2016}.
The radius and the pitch determine the curvature $\tilde{\kappa}$ and  torsion $\tilde{\tau}$  of the helix,
\begin{align}
\tilde{\kappa}=R/L^{2}\ ,\ \ \tilde{\tau}=p\Delta h/(2\pi)/L^{2}\ .
\end{align}
In the following we use  normalized values for those,
given in Eqs.~(\ref{eqn:tau}), $\tau=\tilde{\tau}L$ and $\kappa=\tilde{\kappa}L$, and thus $\kappa=\sqrt{1-\tau^2}$.
(The torsion $\tau$ should not be confused with the Pauli matrices in  orbital space,  $\tau^{}_{j}$).
One easily verifies  that the Hamiltonian is time-reversal symmetric. 

The molecule is attached to two terminals [see Fig. \ref{fig:setupDNA}(b)], such that the total Hamiltonian of the system is
\begin{align}
{\mathcal H} = {\mathcal H}^{}_{\rm mol} + {\mathcal H}^{}_L + {\mathcal H}^{}_R + {\mathcal V} \ .
\end{align}
Here, ${\cal H}^{}_{s}$,  with $s=L,R$,  is the Hamiltonian of the   $s$ lead,
\begin{align}
\mathcal{H}^{}_{s} = -J^{}_0 \sum_{n=1}^{N^{}_{{\rm lead} , s}} c_{s,n+1}^\dagger c^{}_{s,n} +  {\rm H.c.} \ ,
 \label{eq:H_R}
\end{align}
where  $c^{\dagger}_{s,n}$ is an  operator in the spinor representation (\ref{eqn:spinor_twoorb}) with $N_{{\rm lead},s}$ entries, and
$N_{{\rm lead} , s}$ is the number of sites on the $s$ lead, which eventually is assumed to approach infinity.
The tunneling Hamiltonian  connecting the molecule with the terminals reads as
\begin{align}
{\mathcal V} = v c_{1}^\dagger c^{}_{R,1} + v c_{N_{\rm mol}}^\dagger c^{}_{L,1}+  {\rm H.c.}\ ,
\label{eqn:v}
\end{align}
where the tunneling matrix element $v$ is taken to be a real number to preserve the time-reversal symmetry of the entire Hamiltonian. As seen, the right terminal is connected  with the first site on the molecule, and the left terminal with the last one; it is assumed that the tunneling between the molecule and the leads does not mix the orbitals or the spin states.


\subsection{The scattering matrix of the helical junction }
\label{sec:S_matrix_spin_current}


The scattering matrix corresponding  to our  helical junction is an 8$\times$8 matrix, as the scattering waves comprise  four-dimensional spinors [see Eq. (\ref{eqn:spinor_twoorb})]. It is given by the canonical expression
\begin{align}
S = {\bf 1}^{}_8 -2 i \pi W(E)^\dagger {\bm G}(E) W(E) \ .
\label{eqn:S_matrix}
\end{align}
where  ${\bf G}(E)$ is the Green's function at energy $E$ {\em of the entire system}, the molecule and the attached terminals.
Obviously, the Green's function is a matrix of order 4$N^{}_{\rm mol}\times$4$N_{\rm mol}$, of the form
\begin{align}
{\bm G}(E)^{-1} _{}= (E+i \eta) {\bm 1}^{}_{4 N_{\rm mol}}  - {\bm H}^{}_{\rm mol} - {\bm \Sigma} (E) \ ,
\label{GF}
\end{align}
where $\eta$ is a positive infinitesimal.
The Hamiltonian ${\bm H}^{}_{\rm mol}$ is a 4$N^{}_{\rm mol}\times$4$N_{\rm mol}$ matrix derived from ${\cal H}^{}_{\rm mol}$ given in Eq. (\ref{eq:H_mol}),
\begin{align}
{\mathcal H}_{\rm mol} = {\bm c}_{\rm mol}^\dagger {\bm H}^{}_{\rm mol} {\bm c}^{}_{\rm mol}\ ,
\label{Hmath}
\end{align}
where
${\bm c}_{\rm mol}^\dagger = \left[ c_1^\dagger,\cdots,c_{N_{\rm mol}}^\dagger \right]$,
each entry of which is  the spinor in Eq.
(\ref{eqn:spinor_twoorb}).
The self energy ${\bm \Sigma} (E)$ arises from the coupling of the molecule to the leads,
\begin{align}
{\bm \Sigma} = ( \Sigma^{}_L \Pi^{}_1 +\Sigma^{}_R \Pi^{}_{N_{\rm mol}}) \otimes {\bm 1}_4\ ,
\end{align}
where $\Pi^{}_j = {\rm diag} \left( {\bm e}^{}_j \right)$, ${\bm e}^{}_j$ being an $N^{}_{\rm mol}-$component horizontal unit vector whose only nonzero  component is the $j$th entry which is 1.
The self-energy $\Sigma^{}_{s}$, with $s=L,R$   is the $\{1\alpha\sigma,1\alpha\sigma\}$ entry of the lead-$s$ Hamiltonian (\ref{eq:H_R}),
\begin{align}
\Sigma^{}_{s}(E) &= v^2 \left[ (E+i \eta) {\bm 1}^{}_{4 N_{{\rm lead},s}}-{\bm H}^{}_s)^{-1} \right]^{}_{1 \alpha \sigma,1 \alpha \sigma} \nonumber \\ &= (v^2/J_0)(z^{}_{0}/2- i [1-z_{0}^2/4]^{1/2}_{} ) \, ,
\label{Sigmas}
\end{align}
where $z^{}_{0}=(E+i \eta)/J_0$, and ${\bm H}^{}_{s}$ is written
 in terms of ${\bm c}_s^\dagger = \left[ c_{s,1}^\dagger,\cdots,c_{s,N_{{\rm lead}, s} }^\dagger \right]$, as derived from Eq. (\ref{eq:H_R}),
\begin{align}
{\mathcal H}^{}_s = {\bm c}_{s}^\dagger {\bm H}_{s} {\bm c}^{}_{s}\ .
\end{align}
[Equation (\ref{Sigmas}) is the well-known result for the self energy due to coupling with a semi infinite one-dimensional chain.]
Finally, the hybridization with the leads, $W(E)$ [see Eq.  (\ref{eqn:S_matrix})] is a 8$\times$4$N^{}_{\rm mol}$ matrix
\begin{align}
W(E)^\dagger_{}= (w^\dagger_{} \otimes {\bm 1}^{}_4) v {\bm \rho}(E)^{1/2}\ ,
\end{align}
 where
\begin{align}
w^\dagger = \left[ \begin{array}{c}{\bm e}^{}_1  \\  {\bm e}^{}_{N_{\rm mol}} \end{array} \right]  = \left[ \begin{array}{ccccc} 1 & 0 & \cdots & 0 & 0 \\ 0 & 0 & \cdots & 0 & 1 \end{array} \right] \, , \label{eqn:w_}
\end{align}
is a 2$\times$$N^{}_{\rm mol}$ matrix.
The matrix of the density of states in the  leads is related to the self-energy,  $-2 \pi i v^2 {\bm \rho} (E) = {\bm \Sigma}(E)-{\bm \Sigma}(E)^\dagger$.

The scattering matrix is self dual.
This can be verified by noting that the
self-energy matrix ${\bm \Sigma}$ is diagonal (and thus it is obviously self dual) and the Hamiltonian (\ref{Hmath}) is self dual,
${\bm H}^{}_{\rm mol}= \left( {\bm 1}^{}_{N_{\rm mol}} \otimes {\tau}^{}_0 \otimes {\sigma}^{}_y \right) {\bm H}_{\rm mol} ^{T} \left( {\bm 1}^{}_{N_{\rm mol}} \otimes {\tau}_0 \otimes {\sigma}^{}_y \right)$,
which is a consequence of the self duality of spin-orbit interaction.


\subsection{Numerical results for the spin polarization}
\label{sec:numerical_result}

As explained in Sec. \ref{sec:spin_filter},  our two-terminal, two-channel junction  allows for spin polarization solely along $z$. That polarization   requires explicit results only for
the conductances $G_{z; ss}^{}$ and $G^{}_{0;ss}$  [the latter is the charge conductance, see Eqs. (\ref{eqn:Gj_})]. Exploiting our expressions for the scattering matrix  in Sec. \ref{sec:S_matrix_spin_current}, we have computed numerically these conductances and extracted from them the  spin polarization as a function of the energy $E$,
for right-handed chirality, i.e., $p=1$ in Eq. (\ref{tphi}). All energies are measured in units of $J$, and the bandwidth in the leads is assumed to be equal to that of the molecule, $J=J^{}_{0}$. (We have found that when this is not the case, the spin polarization is suppressed.).  The strength of the spin-orbit coupling  is chosen  to be $\Delta^{}_{\rm so}/J=0.4$. This estimate is based on a bandwidth \cite{Gutierrez2012} of
$4J \sim 120$ meV  and the intra-atomic  spin-orbit coupling energy in carbon nanotubes  \cite{HuertasHernandoPRB2006}
$\Delta^{}_{\rm so} \sim 12$ meV.

\begin{figure}[ht]
\includegraphics[width=0.8 \columnwidth]{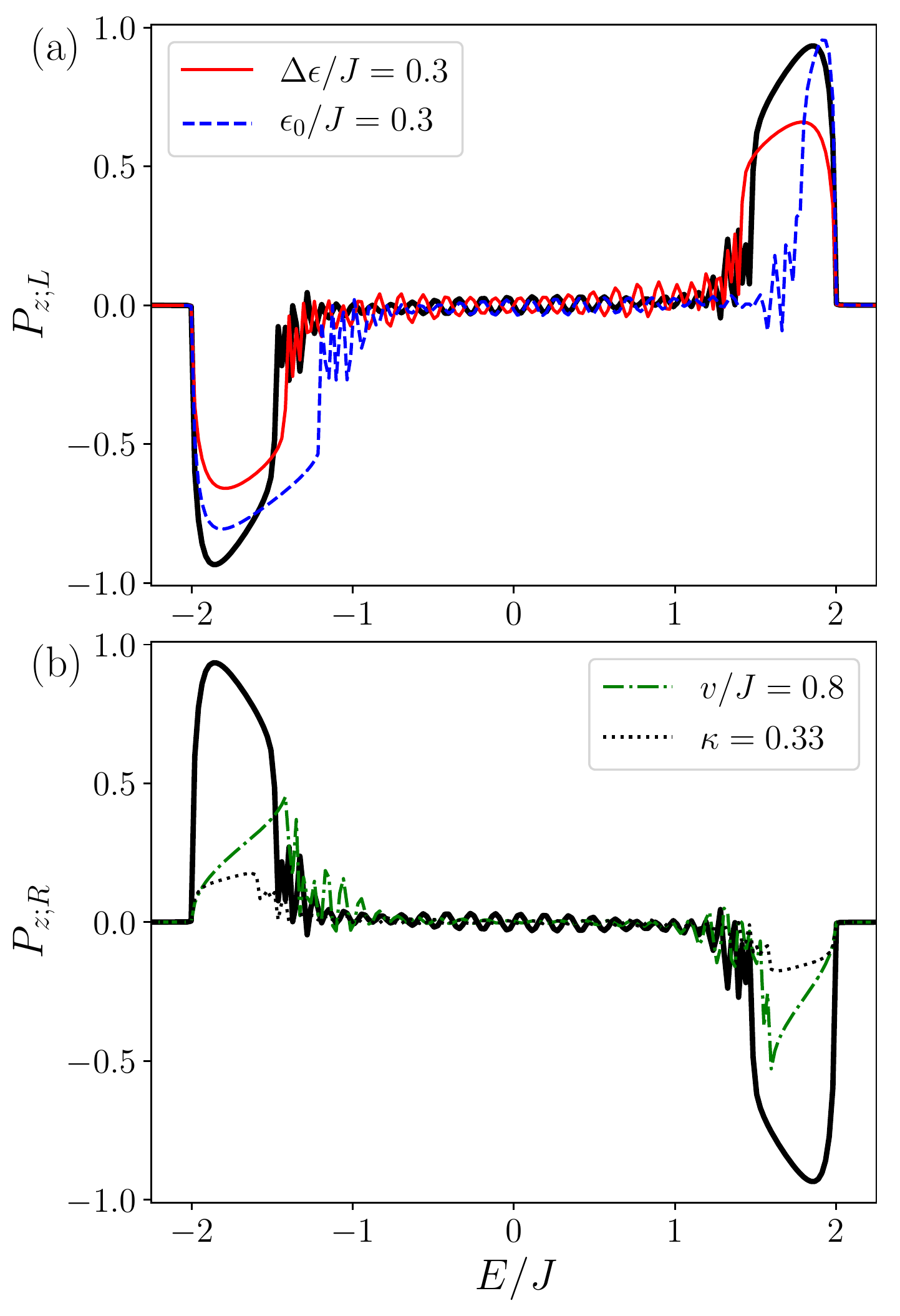}
\caption{
(Color online) Energy dependence of the spin-polarization factor along $z$ for a right-handed helical molecule of length $N_{\rm mol}=20$, in which the number of sites in the unit cells is $N =10$. The spin-orbit coupling  is  $\Delta^{}_{\rm so}/J=0.4$. The spin polarization in the left lead,
$P_{z;L}$,  is shown in (a) and  the one for the right lead,  $P_{z;R}$, is shown in (b).
The thick solid lines are for an  optimal   configuration:  the on-site energies are all identical for the two orbitals, and  the torsion $\tau$ vanishes, $\Delta \epsilon/J=\epsilon_0/J=\tau=0$. The tunnel coupling of the molecule with the leads  is $v/J=1.2$.
The other curves present the polarization for deviations away from  the optimal situation, as marked in the legends. }
\label{fig:PvsEparam}
\end{figure}

The two panels in Fig.~\ref{fig:PvsEparam} show the  $z$ component of the spin polarization factor, Eq.  (\ref{eqn:P}),  measured at the left lead (a) and at the right one (b).
The thick solid lines correspond to an  optimal configuration explained in the  caption.
As seen,  the  spin polarization is large  
in the range $2J-2\Delta^{}_{\rm so}<|E|<2J$,
and the spin polarization factors have opposite signs in the two leads. Since the chirality of the molecule is reversed  when  observed from the opposite lead, this fact implies that the chirality determines the direction of the spin polarization.
In other words, the directions of the spin-polarized currents flowing through the left and right leads are opposite. 
This feature, which might be checked experimentally, is apparently specific for the two-terminal CISS effect.
It was also found in a study of a non-CISS system, where spin filtering is established by applying a magnetic field~\cite{Shekhter2018}.

\begin{figure}[ht]
\includegraphics[width=0.8 \columnwidth]{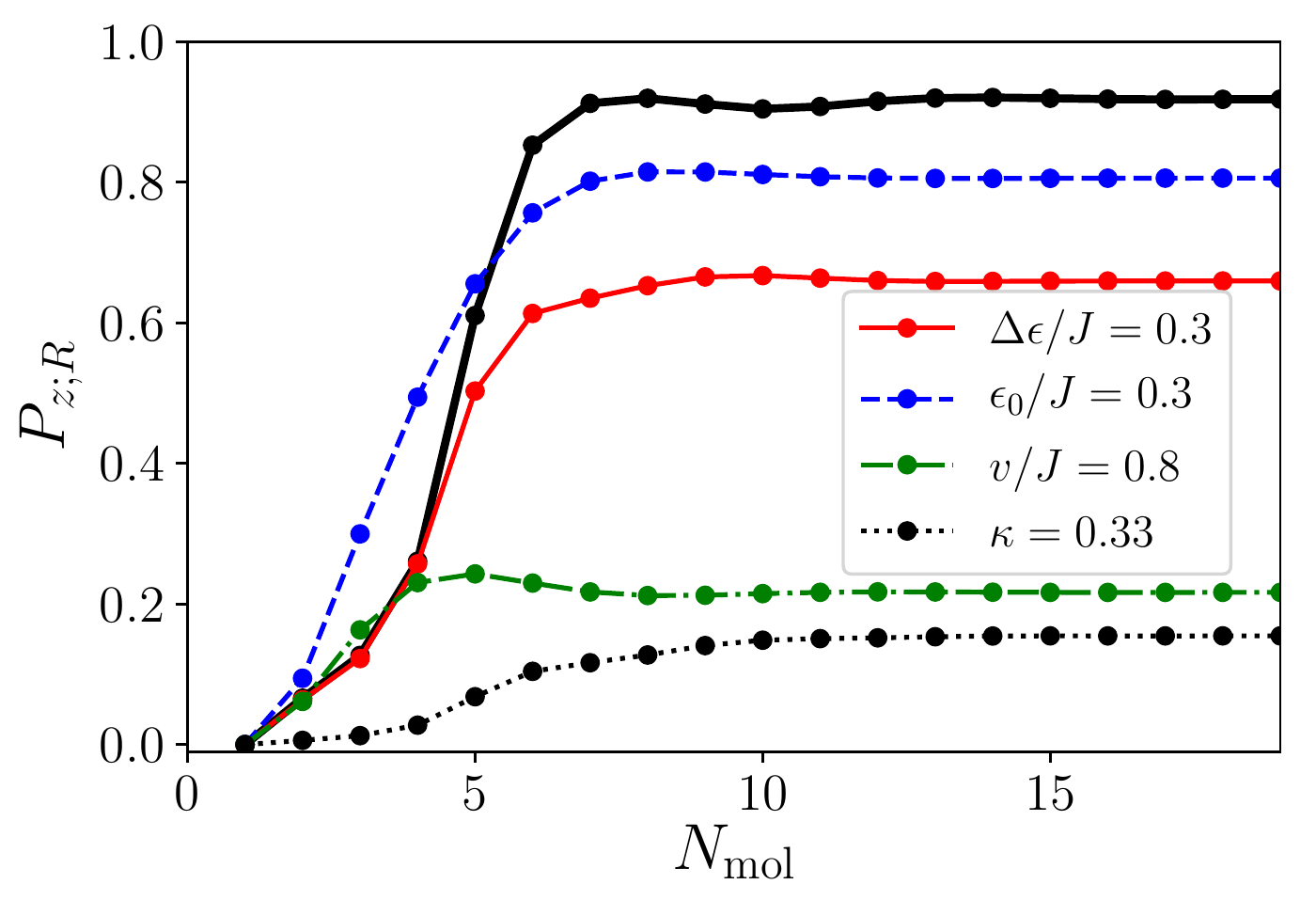}
\caption{(Color online) The length dependence of  $z$ component of the spin  polarization factor in the right lead.
The energy is fixed at $E/J=-1.8$.
Other parameters are as in Fig.~\ref{fig:PvsEparam}.
}
\label{fig:PvsNparam}
\end{figure}

Figure \ref{fig:PvsNparam} shows the length dependence of the spin-polarization factor for a fixed energy,  $E/J=-1.8$, where the positive spin polarization factor  in Fig.~\ref{fig:PvsEparam}(b) is the largest.
The spin polarization increases  as the molecule lengthens, as was found in   a previous study~\cite{Matityahu2016}, in accordance   with experiments~\cite{Goehler2011,Xie2011}. Importantly, the spin polarization becomes almost independent of the  length of the molecule, once the latter exceeds the length of the unit cell.

Our numerical studies show that a finite value for the on-site energy $\epsilon^{}_0$ suppresses the spin polarization [Fig.~\ref{fig:PvsEparam}(a)]. Likewise, a finite difference,  $\Delta \epsilon$, between the energies of two orbitals suppresses the spin polarization [Fig. \ref{fig:PvsEparam}(a)]. The spin-filtering effect is  rather sensitive to the value of the
tunnel matrix element $v$ [Fig. \ref{fig:PvsEparam}(b)].
Furthermore,  one notes that a reduction in the curvature
$\kappa$  reduces considerably the spin polarization [Fig.~\ref{fig:PvsEparam}(b)].
In our case, $\Delta h/R =18.1$ (see Ref.~\onlinecite{Sasao2019}), and thus the `normalized' curvature is $\kappa \approx 0.33$.
The spin polarization seems to be sensitive to perturbations which mix left- and right-going waves, such as  interface scattering and scattering between the sub-systems induced by a finite torsion [see Sec. \ref{sec:band_structure} for the definition of the sub systems, in particular, Eq.~(\ref{eqn:H_pm})].

In the next two sections we substantiate these findings in two ways. First, we analyze the band structure of an infinite molecule, and try to relate its characteristics to the appearance of   spin polarization in the transport. Second, we present explicit  expressions for the scattering matrix of the smallest possible molecule, which are used to derive an analytic result for the spin polarization.


\section{Band structure}
\label{sec:band_structure}

The energy spectrum of a closed system,  i.e., the band structure,  helps to access the origin of  the CISS effect~\cite{VarelaPRB2016,Matityahu2016,Medina2015,SierraBioMol2020}.
We derive the spectrum of  the  Hamiltonian   (\ref{eq:H_mol}) assuming
 that  the molecule comprises $M$ unit cells and obeys periodic boundary conditions (the Born-von Karman  conditions~\cite{AshcroftMermin}).   It is useful  to change the site index  $n$  to $N m+n$,  where $n$ runs on the sites in the unit cell,  $n=1,2, \cdots, N$ and  $m$ numbers the  unit cells.
The periodic boundary condition is then
\begin{align}
c^{}_{n+NM}=c^{}_n \ .
\label{eqn:periodic_bc}
\end{align}
The energy spectrum is  discussed for a very long molecule, i.e., for $M\rightarrow\infty$.

\begin{figure}[ht]
\includegraphics[width=0.9 \columnwidth]{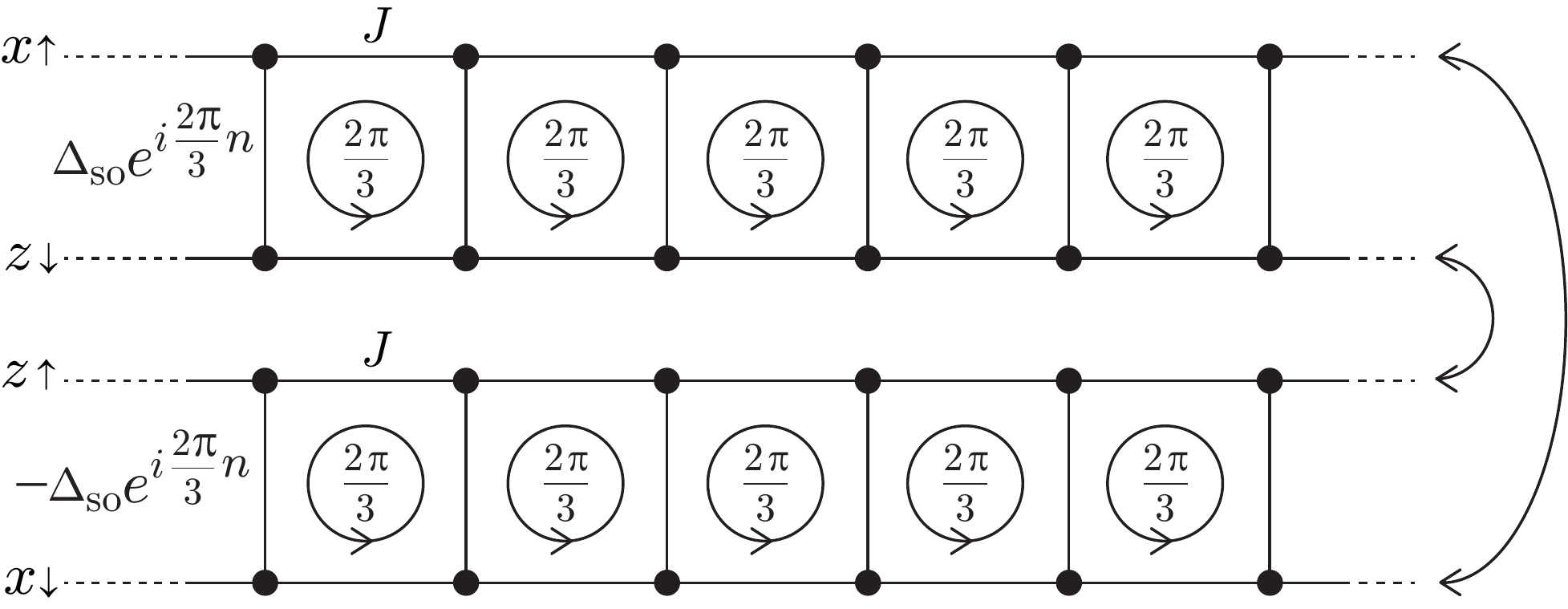}
\caption{
Ladders threaded by  a  fractional flux induced by the helical structure, for  three sites in the unit cell.
The double arrows connect the time-reversed partners.
The vertical lines represents the tunneling amplitudes,
$\pm \Delta^{}_{\rm so} \exp[i 2 \pi n/3]$, connecting the $\uparrow$ and $\downarrow$ spin states (that belong to different orbitals) at the $n$th rung.
The site index $n$ increases from left to right.
 }
\label{fig:ladders}
\end{figure}

The Hamiltonian of the molecule,  Eq.  (\ref{eq:H_mol}), is expressed in terms of the spinors given in Eq. (\ref{eqn:spinor_twoorb}). That scheme  was used for studying numerically the scattering matrix of a single molecule (Sec.  \ref{sec:numerical_result}). However, in order to use all symmetries  in the calculation of the band structure,  it is expedient to reorganize the spinors such that the chain separates
 into two sub systems. Accordingly, we define
\begin{align}
c^{\dagger}_{n;+} = \left[ \begin{array}{cc} c^{\dagger}_{n;x \uparrow} & c^{\dagger}_{n;z \downarrow} \end{array} \right],
\ \ c^{\dagger}_{n;-} = \left[ \begin{array}{cc} c^{\dagger}_{n;z \uparrow} & c^{\dagger}_{n;x \downarrow} \end{array} \right] \, ,
\label{eqn:ann_FB}
\end{align}
which are time-reversed partners \cite{Bernevig2013},
$\Theta c^{\dagger}_{n;\pm \sigma} \Theta^{-1}= \sigma c^{\dagger}_{n;\mp \bar{\sigma} }$.
Note that the Pauli matrices act in the pseudo-spin space, where up and down spins reside on different orbitals.  The Hamiltonian (\ref{eq:H_mol}) becomes
\begin{align}
{\mathcal H}^{}_{\rm mol} = {\mathcal H}^{}_++{\mathcal H}^{}_-+{\mathcal H}^{}_{+-}\ ,
\label{HMOL}
\end{align}
where ${\mathcal H}^{}_{+}$ and   ${\mathcal H}^{}_{-}$  correspond to two sub systems,
\begin{align}
{\mathcal H}^{}_\pm =&\sum_{n=1}^{NM} \Big ( [- J c_{n+1;\pm}^\dagger c^{}_{n;\pm} +{\rm H.c.}  ]\pm \Delta \epsilon \,  c_{n;\pm}^\dagger \sigma^{}_z c^{}_{n;\pm} \nonumber \\ & \pm  p \kappa \Delta^{}_{\rm so} \, c_{n;\pm}^\dagger \left[\begin{array}{cccc} 0 & e^{-i p \phi^{}_n} \\ e^{i p \phi^{}_n} & 0 \end{array} \right] c^{}_{n;\pm}\Big )\ ,
\label{eqn:H_pm}
\end{align}
are coupled together by ${\mathcal H}^{}_{+-}$,
\begin{align}
{\mathcal H}^{}_{+-} =& \sum_{n=1}^{MN}  i |\tau| \Delta^{}_{\rm so} ( c_{n;+}^\dagger c^{}_{n;-} - c_{n;-}^\dagger c^{}_{n;+} ) \ .
\label{eqn:H_pmc}
\end{align}
(The band center   $\epsilon^{}_{0}$  is chosen as the energy reference, $\epsilon^{}_{0}=0$.)
The two Hamiltonians ${\mathcal H}_+$ and ${\mathcal H}_-$ are time-reversed partners, i.e.,  $\Theta {\mathcal H}_\pm \Theta^{-1} = {\mathcal H}_\mp$.
Each of those describes a ladder with a  fractional flux resulting from the helical structure,  that threads each window (see Fig. \ref{fig:ladders}).
When the `normalized' torsion $\tau$ vanishes (which consequently increases  the  spin-orbit coupling, since the `normalized' curvature,  $\kappa=\sqrt{1-\tau^2}$, is then increased),  ${\cal H}^{}_{+-}=0$ and  the two sub systems are decoupled.

We next apply the Bloch theorem to our periodic system, using
 the discrete Fourier expansion
\begin{align}
c^{}_{n ; \pm} = \frac{1}{\sqrt{MN}} \sum_{\ell=0}^{MN-1} e^{i k^{}_\ell n/N} c^{}_{k^{}_\ell ; \pm}\ ,\ \
k^{}_\ell=\frac{2 \pi \ell}{M}\ ,
\label{eqn:F_T}
\end{align}
which obeys the periodic boundary condition (\ref{eqn:periodic_bc}).
Inserting the expansion (\ref{eqn:F_T}) into the spin-orbit term of the Hamiltonian gives
\begin{align}
&  \sum_{n=1}^{NM}c_{n;\pm}^\dagger \left[\begin{array}{cccc} 0 & e^{-i p \phi^{}_n} \\ e^{i p \phi^{}_n} & 0 \end{array} \right] c^{}_{n;\pm}\nonumber\\
&=\sum_{\ell=0}^{NM-1}\left [\begin{array}{cc} 0&  c^{\dagger}_{k^{}_{\ell};\pm \uparrow}
c^{}_{k^{}_{\ell}+2\pi p;\pm\downarrow}\\
c^{\dagger}_{k^{}_{\ell}+2\pi p;\pm\downarrow}
c^{}_{k^{}_{\ell};\pm \uparrow}&\  0
\end{array}\right ]\ .
\label{eqn:spin_oribit_momentum_flip}
\end{align}
It follows that it is useful to define new operators, $a_{k^{}_{\ell};\pm}$, such that
\begin{align}
a^{\dagger}_{k^{}_{\ell};+}&=\left [\begin{array}{cc}c^{\dagger}_{k^{}_{\ell};x\uparrow}&  c^{\dagger}_{k^{}_{\ell}+2\pi p;z\downarrow}\end{array}\right ]\ ,\nonumber\\
a^{\dagger}_{k^{}_{\ell};-}&=\left [\begin{array}{cc}c^{\dagger}_{k^{}_{\ell};z\uparrow}&  c^{\dagger}_{k^{}_{\ell}+2\pi p;x\downarrow}\end{array}\right ]\ ,
\label{aopn}
\end{align}
to obtain
\begin{align}
{\mathcal H}^{}_\pm &= \sum_{\ell=0}^{MN-1} a_{k^{}_\ell ;\pm}^\dagger {\mathcal H}^{}_\pm(k^{}_\ell) a_{k^{}_\ell ;\pm} \ ,
\end{align}
with
\begin{align}
{\mathcal H}^{}_\pm(k^{}_\ell) &= \left[ \begin{array}{cc} E(k^{}_\ell) \pm \Delta \epsilon & \pm p \kappa \Delta^{}_{\rm so} \\ \pm p \kappa \Delta^{}_{\rm so} & E(k^{}_\ell+2 \pi p) \mp \Delta \epsilon \end{array} \right] \ , \label{eqn:h_pm_block}
\end{align}
and
\begin{align}
E(k^{}_\ell)=-2J \cos (k^{}_\ell /N)\ .
\label{Ek}
\end{align}
 The coupling Hamiltonian connecting the two subsystems, in terms of the operators  (\ref{aopn})  is
\begin{align}
{\mathcal H}_{+-} = \sum_{\ell=0}^{MN-1}  i |\tau| \Delta^{}_{\rm so} ( a_{k_\ell;+}^\dagger a_{k_\ell;-} - a_{k_\ell;-}^\dagger a_{k_\ell;+} ) \ ,
\end{align}
and,  as mentioned,  vanishes when $\tau=0$.

\begin{figure}[ht]
\includegraphics[width=1 \columnwidth]{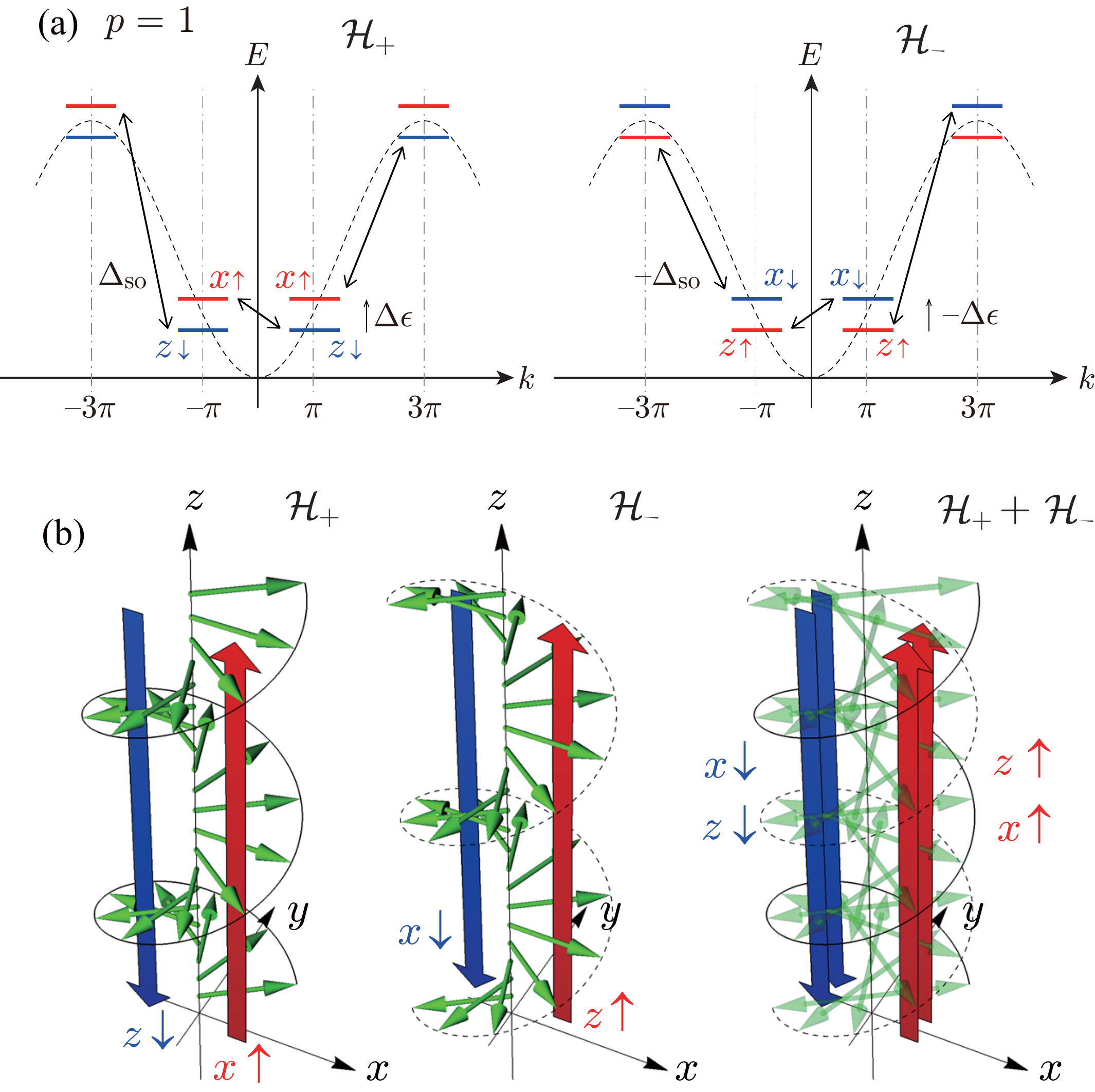}
\caption{
(Color online) (a) States mixed by the effective rotating Zeeman field induced by the spin-orbit interaction ($p=1$ and $N=3$), as indicated by the double arrows.
States at $k=\pi$ and $k=-\pi$ satisfy the condition $E(k)=E(k+2 \pi p)$
($\Delta \epsilon$ is taken to be finite in both panels,  to distinguish $\uparrow$- and $\downarrow$-states.)
The left (right) panel is for ${\mathcal H}_+$ (${\mathcal H}_-$).
Here $\tau=0$ so that the two subsystems are uncoupled.
(b) Cartoon pictures explaining the origin of $\uparrow$- and $\downarrow$-spin polarized states propagating in the opposite directions ($N=10$).
The left and middle panels are for the two sub systems, ${\mathcal H}_+$ and ${\mathcal H}_-$ respectively.
The right panel is for the entire system, ${\mathcal H}_+ + {\mathcal H}_-$.
The coordinates axes $x$, $y$, and $z$ refer to the  spin directions.
The site index $n$ in Eq.~(\ref{eqn:H_pm}) increases along the $z$ axis. }
\label{fig:bandrabi}
\end{figure}

Notice that in the Bloch Hamiltonian (\ref{eqn:h_pm_block}), the wave number of the down spin is shifted to $k_\ell+2 \pi p$ [see Eq. (\ref{aopn})].
In the absence of the spin-orbit interaction, this  corresponds just to a shift in the band index, which (as explained below)   does not entail any physical consequences. However, when $\Delta^{}_{\rm so}\neq 0$  this chirality-dependent additional momentum is crucial.  In fact, the effect of the spin-orbit interaction is equivalent to that of an effective  Zeeman field which rotates  in the $x$-$y$ plane and which  causes transitions, between the states
$| k, x \uparrow \rangle $ and $ |k + 2 \pi p  , z \downarrow \rangle$ for ${\mathcal H}_+$,
and likewise  between the states
$| k, z \uparrow \rangle$
and
$|k + 2 \pi p , x \downarrow \rangle$ for ${\mathcal H}_-$.
This scattering between the two states is dominant when they are energetically degenerate $E(k)=E(k+2 \pi p)$.
For $\Delta \epsilon=0$, this condition is realized for $k=\pi (N j-p)$, where $j$ is an integer.
The two panels in Fig.~\ref{fig:bandrabi}(a) show the states that are mixed when this condition is realized for a right-handed helix,  $p=1$.
The left-going $\uparrow$-spin state with $k=-\pi$ and the right-going $\downarrow$-spin state with $k = \pi$ are mixed considerably due to the effective rotating Zeeman field.
On the other hand, the right-going $\uparrow$-spin state with $k=\pi$ and the left-going $\downarrow$-spin state with $k=-\pi$ are less affected and thus propagate through the helix.
The change of the chirality from right-handedness to  left-handedness, reverses the direction of this propagation.

Figure~\ref{fig:bandrabi}(b)  presents cartoon pictures, meant to explain intuitively  the origin of $\uparrow$ and $\downarrow$ spin-polarized states propagating in opposite directions.
In both sub systems the effective Zeeman fields  rotate around the $z$ axis in the right-hand direction (left and middle panels).
However, the directions of the effective fields for ${\mathcal H}_+$ and ${\mathcal H}_-$ are opposite
(as marked by the green arrows in the left and middle panels) and they cancel one another (right panel).
Since in each sub system the rotating effective field induces an $\uparrow$-spin state propagating in one direction
(along the chain in the panels) and a $\downarrow$-spin state propagating in the other direction
(thick red and blue arrows in the left and middle panels),
 two spin-polarized states are realized without breaking time reversal symmetry (right panel).

For $\tau=0$ and $\kappa=1$ the Hamiltonians ${\cal H}^{}_{\pm}$ given in Eq.
(\ref{eqn:h_pm_block}) are easily diagonalized.
Denoting the band index by $q$,
$q=1,2,\cdots,N$,
the energy takes the form $E_q(k)\equiv E(k+2 \pi q)$. Each of these (uncoupled) Hamiltonians  can be written as
\begin{align}
{\mathcal H}^{}_{\pm}(k) &= \varepsilon^{}_{c}(k,q)\nonumber\\
&+\sqrt{ [\varepsilon^{}_{d}(k,q) \pm \Delta \epsilon]^2 + \Delta^{2}_{\rm so} } \hat{\bm n}^{}_\pm(k,q) \cdot {\bm \sigma}\ ,
\end{align}
where
\begin{align}
\varepsilon^{}_{c}(k,q) &= [E^{}_q(k) + E^{}_{q+1}(k)]/2\ ,\nonumber\\
\varepsilon^{}_{d}(k,q) &= [E^{}_q(k) - E^{}_{q+1}(k)]/2\ ,
\end{align}
and
\begin{align}
\hat{\bm n}^{}_{\pm}(k,q)=\{\sin[\theta^{}_{\pm}(k,q)],0,\cos[\theta^{}_{\pm}(k,q)]\}\ .
\end{align}
The angles $\theta^{}_{\pm}(k,q)$ are the tilting angles of the pseudo-spin away from the $z$ axis, caused by the spin-orbit interaction within each ladder,
\begin{align}
\theta^{}_{\pm}(k,q)={\rm arctan}\{\pm\Delta^{}_{\rm so}/[\varepsilon^{}_{d}(k,q)\pm\Delta\epsilon]\}\  .
\end{align}
The eigenvalues of the matrix
 $\hat{\bm n}^{}_\pm(k,q)\cdot {\bm \sigma}$ are $\beta\equiv\pm 1$ and  the eigen  energies are
 $E^{}_{\pm, \beta}(k,q) = \varepsilon^{}_{c}(k,q) + \beta [(\varepsilon^{}_{-}(k,q) \pm \Delta \epsilon)^2 + \Delta^{2}_{\rm so} ]^{1/2}$,
with $E^{}_{+,\beta}(-k,N-1-q)=E^{}_{-,\beta}(k,q)$.
The $z$ component of the  quantum average of the spin is given by
$\langle \beta; \hat{\bm n}^{}_\pm(k,q) | \sigma^{}_z |\beta; \hat{\bm n}^{}_\pm(k,q) \rangle = \beta \cos[ \theta_\pm(k,q)]$,
where $| \hat{\bm n}^{}_\pm(k,q) \rangle$ is the  eigen ket of ${\mathcal H}^{}_{\pm}(k,q)$, and
$ \cos [\theta^{}_+(-k,N-1-q)]= \cos[ \theta^{}_-(k,q)]$.

\begin{figure}[ht]
\includegraphics[width=1 \columnwidth]{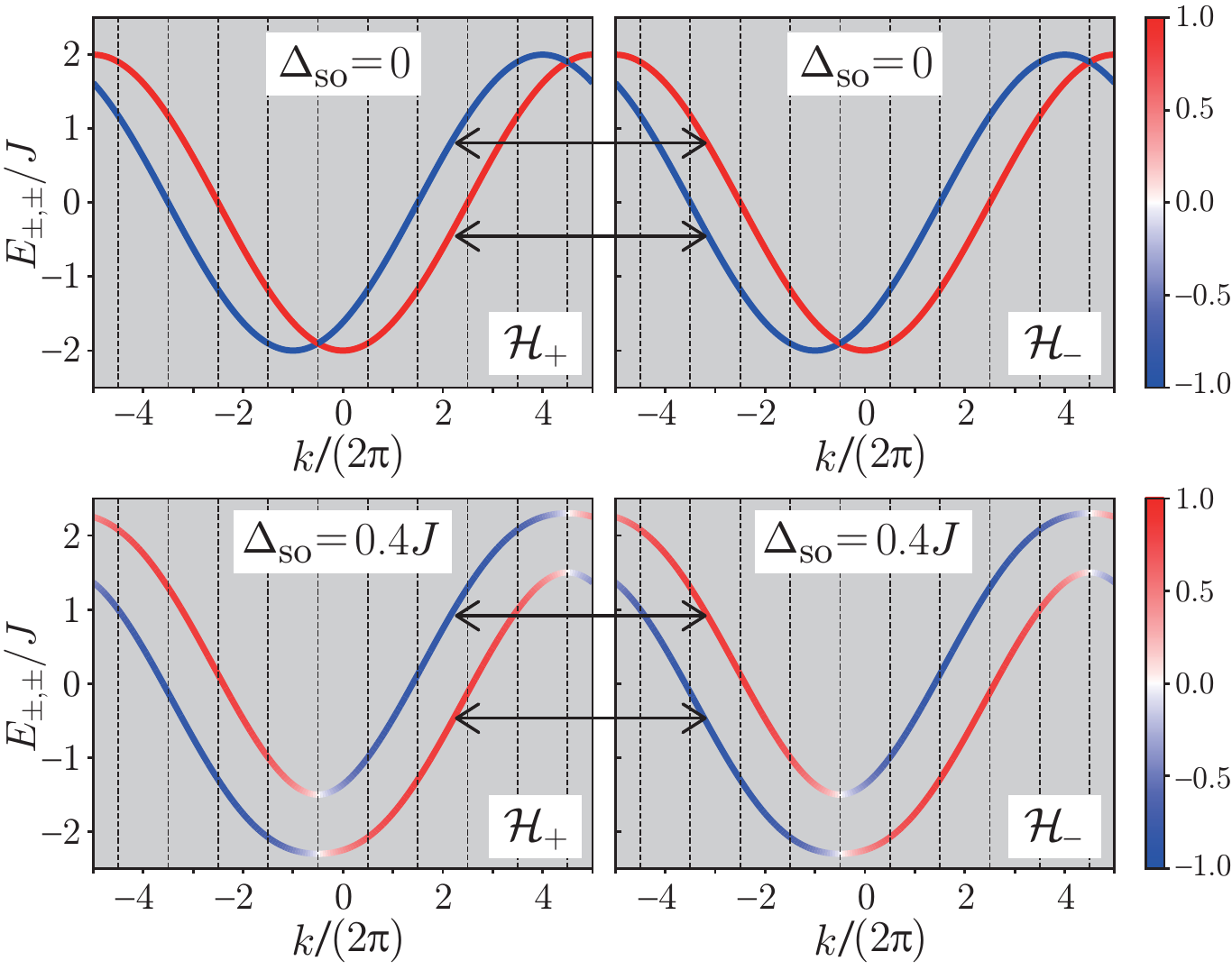}
\caption{
(Color online) Two energy bands  in the extended-zone scheme for 10 sites in the unit cell and right-handed chirality,  calculated for $\tau=0$, i.e., when the two sub systems described by ${\cal H}^{}_{+}$ and ${\cal H}^{}_{-}$ are decoupled.
The color scheme indicates the $z$ component of the average spin (red for $\uparrow$ spins and  blue for $\downarrow$ spins, see the color bar), and the double arrows connect time-reversed partners.
The top (bottom)  panels are for zero (finite) spin-orbit coupling, marked in the figures.  In all plots the on-site energies vanish for both orbitals.
In the absence of the spin-orbit interaction, the $\downarrow$-spin band is shifted by $-2\pi$ with respect to the $\uparrow$-spin band. A finite value of this coupling (bottom panels) causes avoided crossings.
These  occur at the boundaries of the Brillouin zone [$k/(2 \pi)=-1/2 ,\ 9/2$].
}
\label{fig:band_pm}
\end{figure}

The energy dispersion is presented in Fig. \ref{fig:band_pm}, within the extended-zone scheme: the first Brillouin zone  is in the range $-\pi < k \leq \pi$, while the bands are given in the range  $-\pi N < k \leq \pi N$. The spectrum is calculated for    $\tau=0$,  when the the two sub systems are not coupled,  and $p=1$ for the handedness.
The corresponding spectra are shown in the left  (for ${\cal H}_{+}$) and  right  (for ${\cal H}^{}_{-}$) panels.
Time-reversed pairs of states,  $\Theta a^{}_{k_\ell ; \pm \sigma} \Theta^{-1} _{}= \sigma a^{}_{-k_\ell-2 \pi p ; \mp \bar{\sigma}}$,
are connected by double arrows.
In each panel, the color scheme indicates the $z$ component of the pseudo-spin, $\beta \cos [\theta^{}_\pm(k,q)]$, and
the spin-resolved bands are shifted by $-2\pi$  with respect to one another.
In the top panels, the spin-orbit coupling is set to be zero; in that case $-2 \pi$  only changes the band index in the extended-zone scheme.
When the spin-orbit interaction is active (the bottom panels of Fig. \ref{fig:band_pm}),   there appear avoided crossings at boundaries of the Brillouin zone,  $k=- \pi $ and $k=\ 9 \pi$. As seen,
there are two $\uparrow$-($\downarrow$-) spin-polarized right- (left-) going modes within the energy range $-2 J < E < -2J +2 \Delta_{\rm so}$,
which implies positive spin polarization in the right lead, $P_{z;R}>0$.
On the other hand, in the range $2J- 2\Delta_{\rm so} < E < 2J$, there are two $\downarrow$-($\uparrow$-) spin-polarized right- (left-) going modes, which implies
negative spin polarization in the right lead, $P_{z;R}<0$. This explains
qualitatively  the tendency seen in the numerical result, Fig.~\ref{fig:PvsEparam}(b). (Recall that the polarization is almost length independent once the molecule comprises more than a single unit cell, Fig. \ref{fig:PvsNparam}.)

The avoided crossings in the two bottom  panels of Fig.~\ref{fig:band_pm} result from the interplay between the helical structure and the spin-orbit interaction.
Such an interplay has been pointed out before for the Rashba-type SOI, where it appeared as a cutoff of the period of oscillation of the mechanical torque (which is equivalent to   spin current) as a function of the length of molecule~\cite{Sasao2019}.

It is worthwhile to emphasize that the separation of the molecule into two  sub systems described by ${\cal H}^{}_{\pm}$,   realized when  torsion $\tau$ vanishes (and is beneficial for increasing the    spin-orbit coupling), implies a reflection matrix as the one given by Eq.~(\ref{eqn:rm_2c}).
This is further detailed in Sec. \ref{sec:Analytical_S_matrix}.
For zero torsion, an electron moving on the periodic chain encircles the spiral curved line of the molecule [see Eqs. (\ref{eq:H_mol}) and (\ref{tphi})]:
the right- (left-) going electron
encircles the path in the
anti clockwise (clockwise) sense, propagating towards the  positive (negative) direction of the curved helix ({\it cf.} Fig.~\ref{fig:band_pm}), drawn for $p=1$. In this respect,
the directions of spin and  propagation are parallel [in the energy range $- 2 J<E<-2(J-\Delta_{\rm so})$] or anti-parallel [for $2(J-\Delta_{\rm so}) <E<2J$].
Such states were discussed previously in Ref.~\onlinecite{Medina2015}. 

\section{Analytic results for the scattering matrix}
\label{sec:Analytical_S_matrix}


\subsection{Symmetries of the scattering matrix}
\label{syms}

It is illuminating to study the symmetries of our model Hamiltonian,  as those are reflected in the scattering matrix, in  particular  in the `spin conductance' $G^{}_{j;s\bar{s}}$ for $j\neq 0$, given in
Eq.~(\ref{eqn:g_for_analytic2}).
We first show that
$G_{z; \bar{s} s}$ is independent of the sign of the intra-atomic spin-orbit coupling $\Delta^{}_{\rm so}$.
This feature contradicts our findings in Sec.~\ref{QPC} for spin-filtering in a quantum point contact: in that case  the chirality depends on the sign of the electric field inducing the Rashba interaction
and consequently the spin-polarization factor changes sign with the sign of that coupling.

The sign of the spin-orbit coupling in the Hamiltonian is reversed by transforming the tangent vector
${\bf t}(\phi_n)$, Eq. (\ref{tphi}),
\begin{align}
&(\tau^{}_z \otimes \sigma^{}_0) [\Delta^{}_{\rm so} \tau^{}_y \otimes {\bm t}(\phi^{}_n) \cdot {\bm \sigma}] (\tau^{}_z\otimes \sigma^{}_0) \nonumber\\
&= -\Delta^{}_{\rm so} \tau^{}_y \otimes {\bm t}(\phi^{}_n) \cdot {\bm \sigma}\ ,
\end{align}
that is, the Hamiltonian (\ref{Hmath}) satisfies
\begin{align}
&{\bm H}^{}_{\rm mol}(-\Delta^{}_{\rm so}) \nonumber\\
&= ({\bm 1}_{N^{}_{\rm mol}} \otimes \tau^{}_z \otimes \sigma^{}_0) {\bm H}^{}_{\rm mol}(\Delta^{}_{\rm so}) ({\bm 1}_{N_{\rm mol}} \otimes {\tau}^{}_z \otimes {\sigma}^{}_0)\ .
\end{align}
As the scattering matrix obeys the same symmetry, it follows that
$G_{z; \bar{s} s}(\Delta^{}_{\rm so})=G_{z; \bar{s} s}(-\Delta^{}_{\rm so})$.

The symmetry with respect to the interchange of the left and right leads is also of interest.
The interchange of the two edges of the molecule
 corresponds to a change of the site index
$n \to N^{}_{\rm mol}+1 -n$,
transforming in turn the Hamiltonian (\ref{eq:H_mol}) into
\begin{align}
\tilde{\mathcal H}^{}_{\rm mol} &= \sum_{n=1}^{N^{}_{\rm mol}-1} (-J c_{n+1}^\dagger c^{}_{n} + {\rm H.c.}) + \sum_{n=1}^{N^{}_{\rm mol}} \epsilon^{}_0 c_{n}^\dagger c^{}_{n} \nonumber \\ & + \Delta \epsilon c_{n}^\dagger \tau^{}_z \otimes  \sigma^{}_0 c^{}_{n}  \nonumber \\ & + \Delta_{\rm so} c_{n}^\dagger  \tau^{}_y \otimes
{\bm t}(\phi^{}_{N_{\rm mol}+1}-\phi^{}_n)
\cdot {\bm \sigma} c^{}_{n}
\nonumber \\
&=
( \tau_z \otimes \sigma_x U^{}_{N_{{\rm mol}+1} })
{\mathcal H}^{}_{\rm mol}
( \tau^{}_z \otimes U_{N_{{\rm mol}+1} }^\dagger \sigma^{}_x )
\ ,
\label{eqn:left_right_parity}
\end{align}
where $U_{n}$ is given in Eq. (\ref{UT}).
The scattering matrix is transformed as well,
\begin{align}
\left( {\bm 1}_2 \otimes \tau^{}_z \otimes \sigma^{}_x U^{}_{N_{\rm mol}+1} \right)
S
\left( {\bm 1}_2 \otimes \tau^{}_z \otimes \sigma^{}_x U_{N_{\rm mol}+1}^\dagger \right) \, . \label{eqn:parity}
\end{align}
Equation ~(\ref{eqn:g_for_analytic2}) then implies that the interchange of the left and right leads changes the sign of the spin-polarization factor,
$G^{}_{z; s \bar{s}}(p) = - G^{}_{z; \bar{s} s}(p)$.
Using the relation between the original scattering matrix and the scattering matrix in the pseudo-spin basis (\ref{decS}),
\begin{align}
S = \sum_\pm ({\bm 1}_2 \otimes {q_\pm}^\dagger) S_\pm ({\bm 1}_2 \otimes {q_\pm}) \, ,
\end{align}
where $q_\pm = \left( \Pi_\pm,\Pi_\mp \right)$, and the relation
\begin{align}
\left( \tau_z \otimes \sigma_x U_{N_{\rm mol}+1} \right) q_-^\dagger = q_+^\dagger (i \sigma_y) U_{N_{\rm mol}+1}^\dagger \, ,
\end{align}
it is seen that the analytic expression for the scattering matrix satisfies the relation (\ref{eqn:parity}).


Next consider the transformation that reverses the sign of the chirality index $p$. Since
$
(\tau^{}_x \otimes \sigma^{}_y) [\tau^{}_y \otimes {\bm t}(\phi^{}_n,p) \cdot {\bm \sigma}] (\tau^{}_x \otimes \sigma^{}_y) = \tau^{}_y \otimes {\bm t}(\phi^{}_n,-p) \cdot {\bm \sigma}$,
we find
\begin{align}
{\bm H}^{}_{\rm mol}(-p) = ({\bm 1}_{N^{}_{\rm mol}} \otimes \tau^{}_z \otimes \sigma^{}_y) {\bm H}^{}_{\rm mol}(p) ({\bm 1}_{N_{\rm mol}} \otimes {\tau}^{}_z \otimes {\sigma}^{}_y) .
\label{trH}
\end{align}
Then,  Eq.  ~(\ref{eqn:g_for_analytic2}) (keeping in mind that the scattering matrix obeys the same symmetry) implies that
$G^{}_{z; \bar{s} s}(p) = - G^{}_{z; \bar{s} s}(-p)$.
Since $p$ appears only in conjunction with the spin-orbit coupling $\Delta^{}_{\rm so}$, it follows  that in the absence of this coupling $G_{z; \bar{s} s}=-G_{z; \bar{s} s}=0$,  as anticipated.

However, there is a situation in which the spin-orbit coupling is finite, yet $G_{z; \bar{s} s}=0$. This happens when there are two sites in the unit cell, and hence
$\phi_{n}=\pi n$  which implies  that the $x-$component  of the tangent vector ${\bf t}(\phi_n)$ [Eq. (\ref{tphi})] vanishes. As a result, the Hamiltonian is independent of $\sigma^{}_{x}$. In that case, one may apply  a transformation that flips the spin and the orbital, which has the same form as in Eq. (\ref{trH}), except that
$\sigma^{}_{y}$ there is replaced by $\sigma^{}_{x}$ and the sign of $p$ is unchanged. The Hamiltonian and the scattering matrix are invariant under such a transformation, but $G_{z; \bar{s} s}=-G_{z; \bar{s} s}$, which prevents spin filtering.

\subsection{Analytic expressions}
\label{analyticex}

It follows that  the minimal number of sites in the unit cell, required for spin-filtering,  is $N=3$.  In the remaining part of this section we examine  the scattering matrix of a molecule comprising a single unit cell with three sites.  The calculation is carried out exploiting  the wide-band limit  which assumes  that $J^{}_0 \to \infty$ while  the self energy   $\Sigma_s(E)=-i v^2/J^{}_0\equiv-i \Gamma$ [Eq. (\ref{Sigmas})],  remains  finite.
In that limit,
the Green's function (\ref{GF}) of the entire system  becomes
\begin{align}
{\bm G}(E)^{-1} =& E {\bm 1}_{4 N^{}_{\rm mol}}  - {\bm H}^{}_{\rm mol} + i \pi W W^\dagger_{} \ ,
\label{eqn:s_g_target}
\end{align}
where
\begin{align}
\pi \, W W^\dagger_{} = \Gamma [ {\rm diag}(\{1,0,0\})+{\rm diag}(\{0,0,1\}) ] \otimes {\bm 1}^{}_4\ .
\label{WW}
\end{align}

\begin{widetext}
The Hamiltonian ${\bm H}^{}_{\rm mol}$ is presented in Secs. \ref{sec:model_tohtbm} and \ref{sec:S_matrix_spin_current} in the spinor scheme
Eq. (\ref{eqn:spinor_twoorb}), and in Sec. \ref{sec:band_structure} by the spinor scheme Eq.~(\ref{eqn:ann_FB}).
Within the first scheme, the Hamiltonian for the three-site molecule ($N=3$, $M=1$) is
\begin{align}
{\bm H}_{\rm mol}
= J \left[\begin{array}{cccccc}
0 & - i {\bm t}(\phi^{}_1) \cdot {\bm \sigma} & - \sigma^{}_0 & 0 & 0 & 0 \\
i {\bm t}(\phi^{}_1) \cdot {\bm \sigma} & 0 & 0 & - \sigma^{}_0 & 0 & 0 \\
- \sigma^{}_0 & 0 & 0 & -i {\bm t}(\phi^{}_2) \cdot {\bm \sigma} & - \sigma^{}_0 & 0 \\
0 & - \sigma^{}_0 & i {\bm t}(\phi^{}_2) \cdot {\bm \sigma} & 0 & 0 & - \sigma^{}_0 \\
0 & 0 & - \sigma^{}_0 & 0 & 0 & -i {\bm t}(\phi^{}_3) \cdot {\bm \sigma} \\
0 & 0 & 0 & - \sigma^{}_0 & i {\bm t}(\phi^{}_3) \cdot {\bm \sigma} & 0
\end{array} \right]\ ,
\end{align}
where each entry is a real quaternion number, which ensures that the Hamiltonian is a self-dual matrix \cite{Beenakker1997}.
Our aim in this section is to derive analytically  the spin polarization pertaining to a self-dual Hamiltonian,  for the simplified situation where
the torsion $\tau$ vanishes (and then the curvature parameter,  $\kappa$, is simply 1). In that case, as shown in Sec. \ref{sec:band_structure},
the Hamiltonian separates within the spinor scheme Eq. (\ref{eqn:ann_FB}) into two decoupled Hamiltonians, ${\bm H}^{}_{{\rm mol};\pm}$,  which is rather advantageous  for the algebra.
Choosing in addition
$\Delta^{}_{\rm so}=J$, Eq. (\ref{eqn:H_pm}) [see also Eq. (\ref{Hmath})] yields
\begin{align}
{\bm H}^{}_{{\rm mol},\pm} = -J\left[\begin{array}{ccc}
\pm pU \sigma^{}_x U^\dagger_{} & \sigma^{}_0 & 0  \\
\sigma^{}_0 & \pm pU^2_{} \sigma^{}_x (U^\dagger)^2_{} & \sigma^{}_0 \\
0 & \sigma^{}_0 & \pm pU^3_{} \sigma^{}_x (U^\dagger)^3_{}
\end{array} \right]  \ ,\ \ \ U = \exp [ - i p \pi \sigma_z/3 ] = (\sigma^{}_0 -ip \sqrt{3} \sigma^{}_z)/2\ \ ,
 \label{H3}
\end{align}
with $U^{3}_{}=-\sigma^{}_{0}$ and $U\sigma^{}_{x}=\sigma^{}_{x}U^{\dagger}_{}$.  The Green's function corresponding to the Hamiltonian (\ref{H3}) is also a block-diagonal matrix,
${\bm G}^{-1} ={\rm diag} \left( {\bm G}_+^{-1},{\bm G}_-^{-1} \right)$;
assuming for simplicity that $\Gamma=J$, and using the notation $z=E/J$,
we find
\begin{align}
{\bm G}_{\pm}^{-1} =& J U^\dagger \left[ \begin{array}{ccc} g_\pm^{-1} &\  \sigma^{}_0 & 0 \\ \sigma^{}_0 &\ \ \  h_\pm^{-1} & \sigma^{}_0 \\ 0 &\  \sigma^{}_0 & \sigma^{}_x g_\pm^{-1} \sigma^{}_x \end{array} \right] U \ ,
\label{GI}
\end{align}
where
\begin{align}
g_\pm^{-1} =& (z+i) \sigma^{}_0 \pm U^\dagger_{} p\sigma^{}_x U \ ,\ \ h_\pm^{-1} = z \sigma^{}_0 \pm p\sigma^{}_x
\ .
\end{align}

The scattering matrix requires the entries 11, 13, 31, and 33 of the inverse matrix (\ref{GI}), because only those sites are connected with the leads \cite{Oreg}. 
These entries correspond to $RR$, $RL$, $LR$, and $LL$, respectively [see Fig. \ref{fig:setupDNA}(b)]. For our simple model, it is
\begin{align}
S^{}_\pm(E) &= {\bm 1}^{}_{4} -2 i J \left[ \begin{array}{cc} \left[{\bm G}^{}_{\pm} \right]^{}_{11} & \left[ {\bm G}^{}_{\pm} \right]_{1 3} \\ \left[ {\bm G}^{}_{\pm} \right]^{}_{3 1} & \left[ {\bm G}^{}_{\pm} \right]^{}_{33} \end{array} \right] 
= 
\left [\begin{array}{cc}
\sigma^{}_{0}-2iU^{\dagger}_{}[g^{}_{\pm }+g^{}_{\pm}D^{}_{\pm}g^{}_{\pm}]U&\ \ \ \
-2iU^{\dagger}g^{}_{\pm }D^{}_{\pm}\sigma^{}_{x}g^{}_{\pm}\sigma^{}_{x}U\\
-2iU^{\dagger}\sigma^{}_{x}g^{}_{\pm}\sigma^{}_{x}D^{}_{\pm}g^{}_{\pm}U
&\ \ \
\sigma^{}_{0}-2iU^{\dagger}[\sigma^{}_{x}g^{}_{\pm }\sigma^{}_{x}+\sigma^{}_{x}g^{}_{\pm}D^{}_{\pm}g^{}_{\pm}\sigma^{}_{x}]U
\end{array}\right ]
\label{decS}
\end{align}
with
$D^{}_{\pm}=[h^{-1}_{\pm}-g^{}_{\pm}-\sigma^{}_{x}g^{}_{\pm}\sigma^{}_{x}]^{-1}_{}$,
which commutes with $\sigma^{}_{x}$.
\end{widetext}

The explicit calculations of the scattering matrix are presented in Appendix \ref{Tech}. 
There, it is found that
\begin{align}
&-2i J[{\bm G}^{}_{\pm} ]^{}_{13}=
 B^{}_{0}\sigma^{}_{0}\pm i[pB^{}_{x}\sigma^{}_{x}+B^{}_{y}\sigma^{}_{y}]+ipB^{}_{z}\sigma^{}_{z}\ ,
\label{G13}
\end{align}
where
\begin{align}
B^{}_{0}&=[2f^{}_{1}(z)+2-iz]/[f^{2}_{1}(z)+f^{2}_{2}(z)]\ ,\nonumber\\
B^{}_{x}&=(1+iz)/[f^{2}_{1}(z)+f^{2}_{2}(z)]\ ,\nonumber\\
B^{}_{y}&=\sqrt{3}(1+iz)/[f^{2}_{1}(z)+f^{2}_{2}(z)]\ ,\nonumber\\
B^{}_{z}&=-\sqrt{3}(2-iz)/[f^{2}_{1}(z)+f^{2}_{2}(z)]\ ,
\label{B}
\end{align}
and
\begin{align}
{\bm 1}^{}_{2} -2 i J[{\bm G}^{}_{\pm} ]^{}_{11}&
=A^{}_{0}\sigma^{}_{0}\pm i(pA^{}_{x}\sigma^{}_{x}+A^{}_{y}\sigma^{}_{y})\ ,
\label{G11}
\end{align}
where
\begin{align}
A^{}_{0}&=1
-2[(1-iz-z^{2})f^{}_{1}(z)\nonumber\\
&+(1-iz)f^{}_{2}(z)]/[f^{2}_{1}(z)+f^{2}_{2}(z)]\ ,\nonumber\\
A^{}_{x}&=-iz f^{}_{1}(z)/[f^{2}_{1}(z)+f^{2}_{2}(z)]\ ,\nonumber\\
A^{}_{y}&=-\sqrt{3}[-izf^{}_{1}(z)+2f^{}_{2}(z)]/[f^{2}_{1}(z)+f^{2}_{2}(z)]\ .
\label{A}
\end{align}
Here we have introduced
\begin{align}
f^{}_{1}(z)&=2(1-iz)^{2}+iz^{3}\ ,\nonumber\\
f^{}_{2}(z)&=2+(1-iz)^{2}\ .
\end{align}
The other two entries of the scattering matrix are derived in a similar way.
Thus, at zero energy ($z=0$) the scattering matrix comprises  real quaternions.

The results derived above pave the way  to obtain an explicit expression for the spin polarization. As seen from Eq. (\ref{eqn:P}), the polarization is
a quotient of two conductances,  given in Eq. (\ref{eqn:g_for_analytic2}),
\begin{align}
G^{}_{j;RL}=\frac{1}{2\pi}{\rm Tr}\{\sum_{\gamma'=\pm}(\Pi^{}_{\gamma '}\sigma^{}_{j}\Pi^{}_{\gamma'})\sum_{\gamma=\pm}
([S^{}_{\gamma}]^{}_{RL}[S^{\dagger}_{\gamma}]^{}_{LR})\}\ ,
\end{align}
where $\Pi^{}_{+}={\rm diag}(1,0)$ and $\Pi^{}_{-}={\rm diag}(0,1)$,
and the trace is carried out in the pseudo-spin space. Put differently, the conductances corresponding to the two sub systems are added together.
The first factor in the trace implies that in our model  only $G^{}_{0,RL}$ and
$G^{}_{z,RL}$ differ from zero. Using Eqs. (\ref{G13}) and (\ref{B}) one finds
\begin{align}
G^{}_{z;RL}=48 \sqrt{3}p z/(2\pi|f^{2}_{1}(z)+f^{2}_{2}(z)|^{2})\ .
\end{align}
while Eqs. (\ref{G11}) and (\ref{A}) yield
\begin{align}
G^{}_{0;RL}=\frac{1}{2\pi}\frac{8[9+2z^{2}-z^{4}+|f^{}_{1}(z)|^{2}]}{|f^{2}_{1}(z)+f^{2}_{2}(z)|^{2}}\ .
\end{align}
Consequently,
\begin{align}
P^{}_{z;R}=\frac{6\sqrt{3}p z}{9+2z^{2}-z^{4}+|f^{}_{1}(z)|^{2}}\ ,
\end{align}
This expression for the polarization is in full agreement with the numerical results presented
in Sec. \ref{sec:numerical_result} for large enough
$J_0$, $v/J=\sqrt{J_0/J}=\sqrt{100}$. In particular, it shows that the polarization vanishes at zero energy ($z=E/J=0$) and reverses its sign with that of the chirality parameter $p$. As seen in Sec. \ref{syms}, interchanging the roles of the left and right leads reverses the direction of the spin polarization,
$P_{z;R}=-P_{z;L}$.

\section{Summary}
\label{sec:summary}

We have demonstrated that spin-resolved transport can be achieved in a helix-shaped system described by a time-reversal symmetric Hamiltonian and   connected to two leads.
Whether such a phenomenon is possible in principle has been debated and discussed in the literature for quite some time.
Indeed,  while the Bardarson theorem prevents spin selectivity in a single-channel, or a single sub-band, junctions obeying time-reversal symmetry, this is not the case with such junctions that support more channels, or sub bands.
Focusing on the two-channel case, we show that quite generally, its 8$\times$8 scattering matrix has two pairs of doubly-degenerate transmission eigenvalues (the central point in Bardarson's theorem) but those correspond to pairs of identical spins belonging to different channels (or different sub bands), and hence allow for spin selectivity.
Technically speaking, we find that the scattering matrix of the the two-orbital-channel junction consists of complex quaternions--a property identical to spin selectivity.

We substantiate our scenario by introducing a toy model for a DNA-like molecule, that supports $p$ orbitals with anisotropic intra-atom spin-orbit interactions. Solving numerically the scattering matrix of such a molecule, we obtain the resulting spin polarization, and relate it to the band structure of the molecule when detached from the leads.
Our model can be mapped onto two single-orbital tight-binding chains with  effective rotating Zeeman fields induced by the spin-orbit interaction.
The key feature is that although the effective rotating  fields in the two sub systems possess the same chirality, i.e. left or right handedness, their directions are opposite and they cancel each other in the entire system.
To further affirm  the numerical  results, we consider  a particularly simple version of the toy model, and solve it analytically, obtaining an expression for the spin polarization.

The effective  fields resulting from the spin-orbit interaction induce two spin-polarized states, with $\uparrow$- and
$\downarrow$-spins propagating in opposite directions, without breaking time-reversal symmetry.
Although the scenario we propose yields significant spin polarization  for zero torsion (and a finite torsion spoils the perfect spin polarization),
it may explain the origin of the chirality-induced spin selectivity in certain organic molecules.
\YU{In a recent paper~\cite{Chang2018}, it has been demonstrated quite generally that chiral crystals with spin-orbit coupling host Kramers-Weyl fermions, which cause unconventional transport properties. It would be interesting to analyze the CISS effect from such a general view point.}


From the experimental point of view, perhaps the main feature that we find is the strong dependence of the spin-filtering effect on the energy of the charge carriers, in addition to its dependence on the chirality parameter of the helix-shaped molecule.
The latter  results in an experimentally-accessible property: the directions of the spin polarizations in the left and the right leads are opposite.

\begin{acknowledgments}
We thank Dmitri S. Golubev and Naoki Sasao for valuable discussions.
This work was supported by JSPS KAKENHI Grants No. 17K05575, No. 18KK0385, and No. 20H01827, and was partially supported by the Israel Science Foundation (ISF), by the infrastructure program of Israel Ministry of Science and Technology under Contract No. 3-11173, and  by the Pazy Foundation.
\end{acknowledgments}


\begin{appendix}

\section{Effective quasi-one-dimensional Hamiltonian for a quantum point contact}
\label{sec:effective_hamiltonian_QPC}

The full Hamiltonian of a two-dimensional electron gas confined  to a point contact potential and subjected to the Rashba interaction  is
\begin{align}
{\mathcal H}^{}_{\rm QPC} = \frac{\hat{p}^{2}_x+ \hat{p}^{2}_z}{2 m^{}_e} + U(x;z) + \frac{k_{\rm so}}{m^{}_e} (\sigma^{}_x \hat{p}^{}_z - \sigma^{}_z \hat{p}^{}_x) \ ,
\end{align}
where $\hat{p}^{}_{x(z)}=-i \partial_{x(z)}$ are the components of the momentum.
As stated in the main text, the  wave function is decomposed to be $\varphi(x,z)=\sum_{\alpha'} \psi^{}_{\alpha'}(x) \chi^{}_{\alpha'}(z;x)$.
Multiplying the resulting  Schr\"odinger equation,
${\mathcal H}^{}_{\rm QPC} \varphi(x,z)=E \varphi(x,z)$,
 on both sides  by $\chi_{\alpha}^{\ast}(z;x)$,
and  integrating over $z$, yields
$\sum_{\alpha^\prime=1}^\infty {\mathcal H}^{}_{\alpha,\alpha^\prime} \psi^{}_{\alpha^\prime}(x) = E \psi^{}_\alpha(x)$,
where
\begin{align}
{\mathcal H}^{}_{\alpha,\alpha^\prime}
&= \frac{1}{2m^{}_e}
\sum_{\alpha^{\prime \prime}=1}^{\infty}
\left[ ( \hat{p}^{}_x - k^{}_{\rm so} \sigma^{}_z ) \delta^{}_{\alpha,\alpha^{\prime \prime}}  + {\mathcal A}^{}_{\alpha,\alpha^{\prime \prime}}(x)   \right]
\nonumber \\ & \times
\left[ ( \hat{p}^{}_x - k^{}_{\rm so} \sigma^{}_z ) \delta^{}_{\alpha^{\prime \prime},\alpha^\prime}  + {\mathcal A}^{}_{\alpha^{\prime \prime},\alpha^\prime}(x)   \right]
\nonumber \\
& + E^{}_{\perp, \alpha}(x) \delta^{}_{\alpha,\alpha^\prime} + V^{}_{\alpha, \alpha^\prime} \sigma^{}_x
-
\frac{k_{\rm so}^2}{2m^{}_{e}} \delta^{}_{\alpha,\alpha^\prime}
\, . \label{eqn:Hq1d_exact}
\end{align}
This expression  is still exact~\cite{Ulreich1998}.
Here
\begin{align}
{\mathcal A}^{}_{\alpha,\alpha^{\prime}}(x) = \int dz \chi^{\ast}_{\alpha}(z;x) \hat{p}^{}_x  \chi^{}_{\alpha^{\prime}}(z;x) ={\mathcal A}^{\ast}_{\alpha^{\prime},\alpha}(x)
\ . \nonumber
\end{align}
Equation~(\ref{eqn:Hq1d_exact}) is derived by  exploiting the completeness relation~\cite{Ulreich1998},
\begin{align}
\sum_{\alpha=1}^{\infty} \chi^{}_{\alpha}(z;x) \chi_{\alpha}^{\ast}(z';x)=\delta(z-z') \ ,
\end{align}
which yields
\begin{align}
\hat{p}^{}_x {\mathcal A}^{}_{\alpha,\alpha'} = \int dz \chi^{\ast}_{\alpha}(z;x) \hat{p}^{2}_x  \chi^{}_{\alpha'}(z;x) - \sum_{\alpha^{\prime \prime}=1}^{\infty} {\mathcal A}^{}_{\alpha,\alpha^{\prime \prime}} {\mathcal A}^{}_{\alpha^{\prime \prime},\alpha^{\prime}} \ .
\end{align}
For
$\left| \hat{p}_x \ln \psi_{\alpha} (x) \pm k_{\rm so} \right| \gg \left| {\mathcal A}_{\alpha^{\prime},\alpha^{\prime \prime}} \right|$ the matrix element ${\cal A}^{}_{\alpha\alpha'}$ can be discarded for the relevant channels $\alpha$ and $\alpha'$;
in that case one obtains Eq.~(\ref{eqn:Hq1d}).

When the confining potential is mirror-symmetric, $U(z;x)=U(z;-x)$, the quasi-one-dimensional Hamiltonian (\ref{eqn:Hq1d_exact}) is invariant under the simultaneous reflections  $x \to -x$ and $\hat{p}^{}_x \to -\hat{p}^{}_x$, together with a spin flip, $\sigma^{}_z \to -\sigma^{}_z$ [since ${\mathcal A}_{\alpha,\alpha^\prime}(x) \to -{\mathcal A}_{\alpha,\alpha^\prime}(x)$ as $\chi_\alpha(z;x)=\chi_\alpha(z;-x)$; note that $x$ acts as a parameter in the Schr\"{o}dinger equation ${\mathcal H}^{}_\perp(z;x) \chi^{}_\alpha(z;x) = E^{}_{\perp, \alpha} (x) \chi^{}_\alpha(z;x)$].
As a result, the reflection parts of the scattering matrix are invariant under a swap of   the  lead indices, $L \leftrightarrow R$, and the spin indices, $ \uparrow \leftrightarrow \downarrow$,
\begin{align}
r^{}_{\alpha \sigma,\alpha' \sigma'} = r^{\prime}_{\alpha \bar{\sigma},\alpha' \overline{\sigma}'}  \ .
\label{eqn:r_reflectionsymmetric}
\end{align}
The quasi-one-dimensional Hamiltonian (\ref{eqn:Hq1d_exact}) satisfies
\begin{align}
{\mathcal H}^{}_{\alpha,\alpha'}(-k^{}_{\rm so}) = \sigma^{}_y {\mathcal H}^{}_{\alpha,\alpha'}(k^{}_{\rm so}) \sigma^{}_y \, .
\label{eqn:Hq1d_exact_sym}
\end{align}
This symmetry  is reflected  in the symmetry of the scattering matrix,
\begin{align}
S(-k^{}_{\rm so}) = ({\bm 1}^{}_{2} \otimes {\bm 1}^{}_{N_s} \otimes \sigma^{}_y) S(k^{}_{\rm so}) ({\bm 1}^{}_{2} \otimes {\bm 1}^{}_{N_s} \otimes {\sigma}^{}_y) \ ,
\label{eqn:trS_qpc}
\end{align}
where $N_s=\infty$ is the number of channels
(see Sec.~\ref{sec:Analytical_S_matrix} for  a  discussion of symmetries of the scattering matrix).
It follows  that  reversing the sign of $k^{}_{\rm so}$ would  reverse the signs of the $z$ and  $x$ components of the spin.

\section{Effective Hamiltonian of a two-orbital single-stranded DNA with intra-atomic spin-orbit coupling}
\label{sec:effective_hamiltonian}

The radius vector to a  point on a continuous helix of radius $R$ and pitch $\Delta h$ is conveniently represented by the Frenet-Serret formulae.
For the helix in Fig. \ref{fig:setupDNA}(a),
\begin{align}
{\bm R}(\phi) =\{R \cos ( \phi),R \sin (p \phi), \Delta h \, \phi/(2 \pi)\} \ ,
\label{Rphi}
\end{align}
where $p=1$ ($p=-1$) for a helix twisted in the right-handed (left-handed) sense.
In the Frenet-Serret frame, the tangent ${\bm t}$ (along  the helix), normal ${\bm n}$,  and bi-normal ${\bm b}$ vectors at a point on the helix are
\begin{align}
&{\bm t}(\phi) = \{- \kappa \sin (\phi), p \kappa \cos (\phi), |\tau|\}\ ,\nonumber\\
&{\bm n}(\phi) = \{- \cos (\phi), - p \sin (\phi), 0\} \ ,\nonumber\\
&{\bm b}(\phi) = {\bm t}(\phi) \times {\bm n}(\phi) = \{p |\tau| \sin (\phi), - |\tau| \cos (\phi), p \kappa\} \ ,
\label{tnb}
\end{align}
where the `normalized'  curvature and torsion,   $\kappa$ and   $\tau$,  are
\begin{align}
\kappa =& \frac{R}{\sqrt{R^2+[\Delta h/(2 \pi)]^2}}\equiv \cos (\theta )\ ,\nonumber\\
\tau =& \frac{p \Delta h/(2 \pi)}{\sqrt{R^2+[\Delta h/(2 \pi)]^2}} \equiv p \sin (\theta) \ .
 \label{eqn:tau}
\end{align}
%

The position of the $n$th site in the tight-binding scheme  is specified by the radius vector ${\bm R}(\phi_n)$, where the increment of $\phi$ between neighboring sites is $\Delta\phi=2\pi/N$, and $\phi^{}_n = 2 \pi n/N$.
Using Eq. (\ref{Rphi}), the wave function of  the $p^{}_{\alpha}$ orbital
($\alpha=x,y,z$),
at the
 $n$th site is
$\psi_{\alpha}[{\bm r}-{\bm R}(\phi_n)] \delta_{\sigma',\sigma}=  \langle {\bm r}| n; \alpha,\sigma \rangle$.
The ket vector is expressed in terms of the `bare' creation spinor operator and the vacuum state, i.e.,  $|n; \alpha, \sigma \rangle =\tilde{c}^\dagger_{n; \alpha, \sigma}|0 \rangle$.
The time reversal of the bare annihilation operator is given by 
 $\Theta \tilde{c}^{}_{n;\alpha} \Theta^{-1} = i \sigma^{}_y \tilde{c}^{}_{n;\alpha}$. 
 In terms of the bare operators, the tight-binding Hamiltonian for
the model of a single-stranded DNA  molecule is
\begin{align}
{\mathcal H}^{}_{\rm mol} =&\Big ( \sum_{n=1}^{N_{\rm mol}-1} - \tilde{c}_{n+1}^\dagger {\bm J} \otimes \sigma^{}_0 \tilde{c}^{}_n  + {\rm H.c.} \Big ) + \sum_{n=1}^{N_{\rm mol}} \epsilon^{}_0 \, \tilde{c}_{n}^\dagger \tilde{c}^{}_{n} \nonumber \\
- &  2 \Delta^{}_{\rm so} \, \tilde{c}_n^\dagger {\bf L} \cdot {\bf S} \tilde{c}^{}_n + K_{ {\bm t} } \, \tilde{c}_n^\dagger [ ( {\bm t}(\phi_n) \cdot {\bf L})^2 - {\bm 1}_3 ] \tilde{c}^{}_n  \nonumber \\ &+ \Delta \epsilon \, \tilde{c}_n^\dagger  [ ( {\bm b}(\phi_n) \cdot {\bf L})^2 - ( {\bm n}(\phi_n) \cdot {\bf L})^2 ] \tilde{c}^{}_n \ ,
\label{eqn:original_hamiltonian}
\end{align}
 where
\begin{align}
\tilde{c}^{\dagger}_{n} = \left[ \begin{array}{cccccc} \tilde{c}^{\dagger}_{n;x \uparrow} & \tilde{c}^{\dagger}_{n;x \downarrow} & \tilde{c}^{\dagger}_{n;y \uparrow} & \tilde{c}^{\dagger}_{n;y \downarrow} & \tilde{c}^{\dagger}_{n;z \uparrow} & \tilde{c}^{\dagger}_{n;z \downarrow} \end{array} \right] \ .
\end{align}
 The first term on the right-hand side of Eq.~(\ref{eqn:original_hamiltonian}) describes the tunneling between nearest-neighbor sites, with the
tunneling amplitude
 ${\bm J} $ being a
  $3 \times 3$ matrix  in the orbital space. For simplicity we assume that this matrix is isotropic,  ${\bm J}=J {\bm 1}_3$.
In the second term
$\epsilon_0$  is the on-site potential energy.
The third term on the right-hand side of Eq.  (\ref{eqn:original_hamiltonian}) represents the intra-atomic spin-orbit interaction whose strength is denoted $\Delta^{}_{\rm so}$.
Here
${\bf L}= (L_x,L_y,L_z)$ is the vector of the orbital angular-momentum operators
\begin{align}
L^{}_x =& \left[ \begin{array}{ccc} 0 & 0 & 0 \\ 0 & 0 & -i \\ 0 & i & 0 \end{array} \right],
L^{}_y = \left[ \begin{array}{ccc} 0 & 0 & i \\ 0 & 0 & 0 \\ -i & 0 & 0 \end{array} \right], 
L^{}_z = \left[ \begin{array}{ccc} 0 & -i & 0 \\ i & 0 & 0 \\ 0 & 0 & 0 \end{array} \right]
,
\end{align}
and ${\bf S}={\bm \sigma}/2$ is the  vector of the spin angular-momentum, with ${\bm \sigma}$ being the vector of the Pauli matrices.
The other terms in the Hamiltonian describe {\em orbital anisotropies.}
We assume that all electric fields generated by neighboring atoms are accounted for by the on-site orbital anisotropies. The leading anisotropy (the fourth term in the Hamiltonian) is the one along the spiral axis, i.e., along the
tangential direction ${\bm t}(\phi_n)$;  the corresponding energy $K_{\bm t}$ is assumed to be much larger than the other anisotropies. This assumption may be justified by noting that the wave-function spreading along the spiral axis is strongly affected by the crystal field generated by atoms in the neighboring sites. The last term on the right hand-side of Eq.
 (\ref{eqn:original_hamiltonian}) refers to the other two anisotropies,
with
$\Delta \epsilon$ being the difference between the anisotropy energies along the normal direction ${\bm n}(\phi_n)$ and the bi-normal direction ${\bm b}(\phi_n)$.

It is  convenient to perform a rotation in  real space of the `bare' operators  $\tilde{c}^{}_n$, such that
\begin{align}
c^{}_n= O^{}_n \tilde{c}^{}_n\ ,
\end{align}
 where
 \begin{align}
 O^{}_n=e^{i L^{}_x \theta^{}_{p}} e^{i L^{}_z p \phi^{}_{n}}\ .
 \label{On}
 \end{align}
Here $\theta^{}_{p}=\theta$ for $p=1$  and $\theta_{p}^{}=\pi-\theta$ for $p=-1$. This unitary transformation does not change the time-reversal relation of the annihilation operator, as $\Theta c_n \Theta^{-1} = O^{}_n \Theta \tilde{c}^{}_n \Theta^{-1} = i \sigma^{}_y c^{}_n$.
The orthonormal basis vectors of a local coordinate system are chosen to be
$\{ -{\bm n}(\phi),{\bm t}(\phi),{\bm b}(\phi) \}$, Eqs. (\ref{tnb}).
(Although  not the standard choice of the Frenet-Serret frame, it is a convenient one because for $\phi=0$ and $\theta=0$
the vectors $-{\bm n}$,  $\pm{\bm t}$,  and $\pm{\bm b}$ are along the $x$, $y$,  and $z$ axes for $p=\pm 1$, respectively.)
The inner products in Eq.~(\ref{eqn:original_hamiltonian}) for the Hamiltonian
are then all diagonal,
$-{\bm n}(\phi^{}_n) \cdot {\bf L} = O_n^\dagger L^{}_x O^{}_n$,
${\bm t}(\phi^{}_n) \cdot {\bf L} = O_n^\dagger L^{}_y O^{}_n$,  and
${\bm b}(\phi^{}_n) \cdot {\bf L} = O_n^\dagger L^{}_z O^{}_n$.

By exploiting the relation $L_\alpha^2 = {\bm 1}_3 - |\alpha \rangle \langle \alpha|$, ($\alpha=x,y,z$), we obtain
\begin{align}
\tilde{c}_n^\dagger [{\bm 1}^{}_3 - ( {\bm t}(\phi^{}_n) \cdot {\bf L})^2 ] \tilde{c}^{}_n &= c_n^\dagger |y \rangle \langle y| c^{}_n\ ,\nonumber\\
\tilde{c}_n^\dagger [{\bm 1}^{}_3 - ( {\bm n}(\phi^{}_n) \cdot {\bf L})^2 ] \tilde{c}^{}_n &= c_n^\dagger |x \rangle \langle x| c^{}_n \ ,\nonumber\\
\tilde{c}_n^\dagger [{\bm 1}^{}_3 - ( {\bm b}(\phi^{}_n) \cdot {\bf L})^2 ] \tilde{c}^{}_n &= c_n^\dagger |z \rangle \langle z| c^{}_n \ .
\end{align}
%
Since the spin-orbit coupling conserves the total angular moment,
$[ {\bf L} + {\bf S}, {\bf L} \cdot {\bf S} ]=0$ for each component of the total angular momentum,
the unitary transformation
\begin{align}
U^{}_n=e^{i S^{}_x \theta^{}_{p}}e^{i S^{}_z p \phi^{}_n}\ ,
\label{UT}
\end{align}
 in conjunction with the rotation (\ref{On}) commutes with ${\bm L}\cdot{\bm S}$, i.e., $[U^{}_{n}O^{}_{n},{\bf L}\cdot {\bf S}]=0$. It follows that
\begin{align}
\tilde{c}_n^\dagger  {\bf L} \cdot {\bf S} \tilde{c}^{}_n &= c_n^\dagger O^{}_n \, {\bf L} \cdot {\bf S} \, O_n^\dagger c^{}_n \nonumber\\
& = c_n^\dagger U_n^\dagger U^{}_n O^{}_n \, {\bf L} \cdot {\bf S} \, O_n^\dagger U_n^\dagger U^{}_n c^{}_n
\nonumber \\ & = c_n^\dagger   {\bf L} \cdot U_n^\dagger{\bf S}  U^{}_n c^{}_n \ .
\end{align}
As a result,
the Hamiltonian (\ref{eqn:original_hamiltonian}) is diagonal in the orbital anisotropy terms,
\begin{align}
&{\mathcal H}_{\rm mol}= \sum_n - c_{n+1}^\dagger J {\bm 1}^{}_3 \otimes \sigma^{}_0 c^{}_n + {\rm H.c.} + \epsilon^{}_0 c_n^\dagger c^{}_n  \nonumber \\ & - 2 \Delta^{}_{\rm so} c_n^\dagger \, {\bf L} \cdot (U_n^\dagger {\bf S} U^{}_n) \, c^{}_n  \nonumber\\
&+ \Delta \epsilon c_n^\dagger (|x \rangle \langle x| - |z \rangle \langle z|) c^{}_n  - K^{}_{\bm t} c_n^\dagger |y \rangle \langle y| c^{}_n  \, .
\end{align}
The price to pay is the rotation of spin axis.

In the limit of a strong orbital anisotropy, i.e.,  $K_{\bm t} \to \infty$, only the $p_x$ and $p_z$ orbitals contribute to the transport.
In this limit the effective Hamiltonian acts in the
$p_x$ and $p_z$ orbital space, and the operator vector $c_{n}^{}$ contains only four components, see Eq.~(\ref{eqn:spinor_twoorb}). Introducing the Pauli matrices that act on the orbital degrees of freedom,  ${\bm \tau}_\alpha$, we obtain the effective Hamiltonian,
\begin{align}
{\mathcal H}_{\rm mol}  =&\Big ( \sum_n - c_{n+1}^\dagger J \tau^{}_0 \otimes \sigma^{}_0 c^{}_n + {\rm H.c.} \Big )+ \Delta \epsilon \, c_n^\dagger \tau^{}_z \otimes \sigma^{}_0 c^{}_n \nonumber \\ & -+2 \Delta^{}_{\rm so} c_n^\dagger \, {\bm \tau}^{}_y(U_n^\dagger S^{}_y U^{}_n) c^{}_n + \epsilon^{}_0 c_n^\dagger c^{}_n \ .
\end{align}
The Hamiltonian (\ref{eq:H_mol}) is obtained  upon  using
\begin{align}
2 \Delta^{}_{\rm so}  {\bm \tau}^{}_y (U_n^\dagger S^{}_y U^{}_n)  =
 \Delta^{}_{\rm so}  \tau^{}_y \otimes {\bm t}(\phi_n) \cdot {\bm \sigma}
\ .
\end{align}
Note  that this effective Hamiltonian describes a quasi-one-dimensional wire, since the $p_x$ and $p_z$ orbitals allow for a rotation around the spiral axis, the thick curved line in Fig. \ref{fig:setupDNA}(a).
Therefore, our scheme shares a certain similarity with the  helix-shaped tube model discussed in Refs.~\onlinecite{Michaeli2019} and \onlinecite{Geyer2020}.


\section{Technical details for Sec. \ref{sec:Analytical_S_matrix}}
\label{Tech}

It is expedient to present the entries of the  matrix in Eq. (\ref{decS}) in terms of (complex) quaternions.
This is achieved by noting that
\begin{align}
ig^{}_{\pm}=g^{}_{0}\sigma^{}_{0}
\mp i{\bf g}\cdot\sig\ ,
\end{align}
where
\begin{align}
g^{}_{0}&=(1-iz)/[(1-iz)^{2}+1]\ ,\nonumber\\
{\bf g}&=\{g^{}_{x},g^{}_{y},0\}=(\{p,\sqrt{3},0\}/2
)/[(1-iz)^{2}+1]
\ .
\end{align}
It follows that
\begin{align}
iD^{}_{\pm}=D^{}_{0}\sigma^{}_{0}\pm i{\bf D}\cdot\sig\ ,
\end{align}
where
\begin{align}
D^{}_{0}&=f^{}_{1}(z)[(1-iz)^{2}+1]/f^{2}_{1}(z)+f^{2}_{2}(z)\ ,\nonumber\\
{\bf D}&=\hat{\bf x}p\
f^{}_{2}(z)[(1-iz)^{2}+1]/f^{2}_{1}(z)+f^{2}_{2}(z)
\ ,
\end{align}
with
\begin{align}
f^{}_{1}(z)&=2(1-iz)^{2}+iz^{3}\ ,\nonumber\\
f^{}_{2}(z)&=2+(1-iz)^{2}\ .
\end{align}
For $z=0$  both $ig^{}_{\pm}$
and $iD^{}_{\pm}$
are real quaternions.

The explicit expressions for the
entries of the  matrix in Eq. (\ref{decS})  are straightforward to derive.
Thus, the expression  that appears in upper off diagonal entry is
\begin{align}
&-ig^{}_{\pm}D^{}_{\pm}\sigma^{}_{x}g^{}_{\pm}\sigma^{}_{x}
=\frac{1}{f^{2}_{1}(z)+f^{2}_{2}(z)}\Big ([f^{}_{1}(z)+1-\frac{iz}{2}]\sigma^{}_{0}\nonumber\\
& +i\Big\{\mp p(1+iz)\ ,0\ ,-p\sqrt{3}(1-\frac{iz}{2})\Big \}\cdot\sig\Big )\ .
\end{align}
The transformation $U$ rotates an arbitrary vector ${\bf v}$ in the $x$-$y$ plane
\begin{align}
U^{\dagger}_{}\sig\cdot{\bf v}U
&=\sigma^{}_{x}[-\frac{1}{2}v^{}_{x}+\frac{p\sqrt{3}}{2}v^{}_{y}]\nonumber\\
&+\sigma^{}_{y}[-\frac{p\sqrt{3}}{2}v^{}_{x}-\frac{1}{2}v^{}_{y}]+\sigma^{}_{z}v^{}_{z}\ ,
\label{rotv}
\end{align}
making  the upper off diagonal entry of  the scattering matrix (\ref{decS}) to be
\begin{align}
&-2i J[{\bm G}^{}_{\pm} ]^{}_{13}=
-2iU^{\dagger}_{}g^{}_{\pm}D^{}_{\pm}\sigma^{}_{x}g^{}_{\pm}\sigma^{}_{x}U\ ,
\end{align}
leading to Eqs. (\ref{G13}) and (\ref{B}) in the main text.

Likewise, the expression that determines the upper diagonal entry is
\begin{align}
&-ig^{}_{\pm}-ig^{}_{\pm}D^{}_{\pm}g^{}_{\pm}
=\frac{1}{f^{2}_{1}(z)+f^{2}_{2}(z)}\nonumber\\
&\times\Big (-[
(1-iz-z^{2})f^{}_{1}+(1-iz)f^{}_{2}]\sigma^{}_{0}\nonumber\\
&\pm i\Big\{\frac{p}{2}[-iz f^{}_{1}+3f^{}_{2}]\ ,\frac{\sqrt{3}}{2}[-izf^{}_{1}+f^{}_{2}]\ , 0\Big\}\Big)\ ,
\label{bd}
\end{align}
which leads to
\begin{align}
{\bm 1}^{}_{2} -2 i J[{\bm G}^{}_{\pm} ]^{}_{11}&= \sigma^{}_{0}-2iU^{\dagger}_{}[g^{}_{\pm }+g^{}_{\pm}D^{}_{\pm}g^{}_{\pm}]U
\ .\end{align}
This expression yields Eqs. (\ref{G11}) and (\ref{A}) in the main text.


\end{appendix}

\end{document}